\documentclass[12pt,a4paper]{amsart}

\usepackage[utf8]{inputenc}

\usepackage{xcolor}
\usepackage{graphicx}

\usepackage[none]{hyphenat}
\begin{document}

\title{SuperDARN scalar radar equations}
\author{O.I. Berngardt, K.A. Kutelev, A.P. Potekhin}
\address{Institute of Solar-Terrestrial physics SB RAS, 664033, 126a, Lermontova Str., Irkutsk, Russia, P.O.Box 291, e-mail: berng@iszf.irk.ru}
\maketitle

\begin{abstract}
The quadratic scalar radar equations are obtained for SuperDARN radars that
are suitable for the analysis and interpretation of experimental data.
The paper is based on a unified approach to the obtaining radar equations
for the monostatic and bistatic sounding with use of hamiltonian optics
and ray representation of scalar Green's function and without taking into account the polarization effects. 
The radar equation obtained
is the sum of several terms corresponding to the propagation and scattering
over the different kinds of trajectories, depending on their smoothness
and the possibility of reflection from the ionosphere. It is shown
that the monostatic sounding in the media with significant refraction,
unlike the case of refraction-free media, should be analyzed as a
combination of monostatic and bistatic scattering. This leads to strong
dependence of scattering amplitude on background ionospheric density
due to focusing mechanism and appearance of new (bistatic) areas of effective
scattering with signifficant distortion of the scattered signal spectrum.
Selective properties of the scattering also have been demonstrated. 
\end{abstract}

\section{Introduction}

Investigations of ionosphere and magnetosphere with use of international
network of SuperDARN(Super Dual Auroral Radar Network) radars becomes
one of the basic techniques for analysis of the processes in the high-latitude
ionosphere and magnetosphere today \cite{Ruohoniemi_1988,Ruohoniemi_1989,Greenwald_1995,Chisham_2007,Ruohoniemi_2011}.
Expanding of the network to the mid-latitudes allows to investigate
expanding the polar effects to the mid-latitudes during geomagnetic
disturbances and storms\cite{Baker_2007}. In spite of wide use of
the data the inversion technique for this method - obtaining the parameters
of the ionosphere from the received signals is still under development.
Today taking into account the ionospheric refraction to the scattered
signal characteristics is important theoretical problem \cite{Nasyrov_1991,Uspensky_1994,Uspensky_1994a,Ponomarenko_2009,Gillies_2009,Ponomarenko_2010,Gillies_2011,Spaleta_2015}.

The basis for the backscattering technique, used by the SuperDARN
radars is the radar equation. It relates the shape of the correlation
function of the received signal with spectral density of irregularities
and characteristics of background ionosphere.

Currently the SuperDARN radar equation exists only in very simple
approximations \cite{Schiffler_1996}, which can lead to potential
problems in data interpretation. For example, correct taking into
account the refraction in the evaluation of the basic SuperDARN parameter
- irregularities velocity \cite{Gillies_2011}.

Traditionally, to interpret the SuperDARN measurements there is used
an analogy \cite{Schiffler_1996} with the radar equation in refraction-free
case \cite{Tatarsky_1967,Ishimaru_1999}. Initially, this radar equation
was developed for the case of scatterers that are small compared with
the Fresnel zone radius, and not always valid for irregularities with
the size compared with this radius. In \cite{Berngardt_1999,Berngardt_2000}
there was obtained the radar equation under assumption of smooth spatial changes
of spectral density of irregularities. This refraction-free radar
equation is valid for scattering by large-scale ionospheric irregularities
including irregularities elongated with the Earth magnetic field and
studied by SuperDARN radars. But correct taking into account the refraction
was not done yet.

In the paper we obtained the scalar radar equation for SuperDARN radars
within the first propagation hop, taking into account refraction and
single reflection from the ionosphere.

\section{Initial equations}

\subsection{Ray representation of Green's function}

Traditionally the problems of radiowave propagation and scattering
in smoothly inhomogeneous media are analyzed using Green's function
approach. Radiowave propagation in smoothly inhomogeneous media can
be described using geometrical optics (WKB approximation)\cite{Heading_2013,Kravtsov_1990}.
Geometrooptical approach allows to replace extended source (antenna)
by point source and distribution of the transmitted signal over the
ray exit directions (antenna pattern). The full geometrooptical Green's
function is matrix-valued (see, for example \cite{Kravtsov_1996,Liang_1997}):

\begin{equation}
\begin{array}{c}
\overrightarrow{E}(\overrightarrow{r},\omega)=\widehat{G}(\overrightarrow{R}_{0},\overrightarrow{r},\omega)\overrightarrow{a}(\omega)\\
\widehat{G}(\overrightarrow{R}_{0},\overrightarrow{r},\omega)=\frac{1}{F^{1/2}(\overrightarrow{R}_{0},\overrightarrow{r})}e^{i\frac{\omega}{c}\psi(\overrightarrow{R}_{0},\overrightarrow{r})}\hat{\Gamma}(\overrightarrow{R}_{0},\overrightarrow{r})
\end{array}\label{eq:GreensFunction}
\end{equation}

where $\hat{\Gamma}(\overrightarrow{R}_{0},\overrightarrow{r})$ -
polarization matrix, $F^{1/2}(\overrightarrow{R}_{0},\overrightarrow{r})$
- decrease of the signal with the range, $\psi(\overrightarrow{R}_{0},\overrightarrow{r})$
- geometrooptical eikonal; $\overrightarrow{R}_{0}$ is the position
of the transmitter.

The problem of the formula is impossibility to use it in the tasks
of multipath propagation, when signal has a number of different trajectories
to reach the same scattering point\cite{Kravcov_1980}. To generalize
(\ref{eq:GreensFunction}) for the problem of multipath propagation
lets use the ray coordinate system and analyze the field in a given
point as a result of interference of several geometrooptical rays
with different exit angles. In simpler scalar case (which does not
take into account the polarization of the electromagnetic wave) we
can make the equivalent ray representation for scalar Green's function
$G_{go}(\overrightarrow{R}_{0},\overrightarrow{\Lambda},\omega)$,
that after integrating over the ray coordinates results into standard
scalar Green's function:

\begin{equation}
G(\overrightarrow{R}_{0},\overrightarrow{r}',\omega)=\int G_{go}(\overrightarrow{R}_{0},\overrightarrow{\Lambda},\omega)\delta(\overrightarrow{r}'-\overrightarrow{R}(\overrightarrow{\Lambda}))F(\overrightarrow{\Lambda})d\overrightarrow{\Lambda}\label{eq:GreensFunctionRays}
\end{equation}

Here $G_{go}(\overrightarrow{R}_{0},\overrightarrow{\Lambda},\omega)$
is Green's function ray representation as a function of ray arguments
$\overrightarrow{\Lambda}$, antenna position $\overrightarrow{R}_{0}$
and frequency of the signal $\omega$; $\overrightarrow{R}(\overrightarrow{\Lambda})$
is ray trajectory for given ray coordinates $\overrightarrow{\Lambda}$;

\begin{equation}
F(\overrightarrow{\Lambda})=\left\Vert \frac{\partial R_{i}(\overrightarrow{\Lambda})}{\partial\Lambda_{j}}\right\Vert \label{eq:dR_dL}
\end{equation}

is Jacobian to convert from ray coordinates to spatial coordinates
(and at the same time the square of geometrical divergence);$\overrightarrow{\Lambda}=(\widehat{\Lambda};\Lambda_{\Vert})=(\widehat{\Lambda}_{\xi},\widehat{\Lambda}_{\eta};S)$
are the ray coordinates - two ray exit angles and trajectory group
length; $d\overrightarrow{\Lambda}=d\Lambda_{\Vert}d\widehat{\Lambda}_{\xi}d\widehat{\Lambda}_{\eta}$.

After taking into account antenna pattern $g(\widehat{\Lambda},\omega)$
of the transmitter, the signal in the position of the scatterer $\overrightarrow{r}$
can be calculated (with accuracy of constant multiplier) from the
integral representation (\ref{eq:GreensFunctionRays}), as a superposition
of the fields, propagating over the different trajectories:

\begin{equation}
u(\overrightarrow{r},\omega)=\int a(\omega)g(\widehat{\Lambda},\omega)G_{go}(\overrightarrow{R}_{0},\overrightarrow{\Lambda},\omega)\delta(\overrightarrow{r}-\overrightarrow{R}(\overrightarrow{\Lambda}))F(\overrightarrow{\Lambda})d\overrightarrow{\Lambda}\label{eq:signal_representation_rayGreen}
\end{equation}

where 
\begin{equation}
G_{go}(\overrightarrow{R}_{0},\overrightarrow{\Lambda},\omega)=\frac{1}{F^{1/2}(\overrightarrow{\Lambda})}e^{i\frac{\omega}{c}\psi(\overrightarrow{\Lambda},\omega)}\label{eq:RayRepresentationFormula}
\end{equation}
$\psi(\overrightarrow{\Lambda},\omega)$ is eikonal in ray coordinates;
$\overrightarrow{R}(\overrightarrow{\Lambda})$ is ray trajectory;
$a(\omega)$ - transmitted signal spectrum for frequency $\omega$.

The given representation (\ref{eq:signal_representation_rayGreen})
has the simple physical sense: the signal that comes to the point
$\overrightarrow{r}$ is a superposition of geometrooptical rays,
transmitted from the antenna with different ray directions $\widehat{\Lambda}$
and amplitudes $g$. The correspondence of the representation (\ref{eq:signal_representation_rayGreen})
to the traditional geometrooptical representation for the signal can
be shown by integrating (\ref{eq:signal_representation_rayGreen})
over the $d\overrightarrow{\Lambda}$ and taking into account that
integrating of delta-function under integral produces the transformation
Jakobian (\ref{eq:dR_dL}) and summation over all propagation trajectories
$\nu$ to the given point $\overrightarrow{r}$:

\begin{equation}
\int g(\overrightarrow{\Lambda})\delta(\overrightarrow{r}-\overrightarrow{R}(\overrightarrow{\Lambda}))d\overrightarrow{\Lambda}=\sum_{\nu}\left\Vert \frac{\partial}{\partial\Lambda_{j}}R_{i}(\overrightarrow{\Lambda})\right\Vert _{\nu}^{-1}g(\overrightarrow{R}^{-1}(\overrightarrow{r}))\label{eq:delta-function-definition}
\end{equation}

After integration the signal (\ref{eq:signal_representation_rayGreen})
transforms into:

\begin{equation}
\left\{ \begin{array}{l}
u(\overrightarrow{r},\omega)=\sum\limits _{\nu}\frac{a(\omega)g(\widehat{\Lambda}_{\nu},\omega)}{F^{1/2}(\overrightarrow{\Lambda}_{\nu})}e^{i\frac{\omega}{c}\psi(\overrightarrow{R}(\overrightarrow{\Lambda}_{\nu}))}\\
\overrightarrow{r}=\overrightarrow{R}(\overrightarrow{\Lambda}_{\nu})
\end{array}\right.\label{eq:KravcovMultipathFormula}
\end{equation}

The trajectories $\nu$ are defined by second equation of the system
(\ref{eq:KravcovMultipathFormula}) for any given $\overrightarrow{r}$.
This equation is close to discussed in \cite{Kravcov_1980}.

\subsection{Ray equations: Hamilton representation}

The eikonal $\psi(\overrightarrow{\Lambda})$ is defined as the integral
over the curvilinear trajectory, definable as a propagation in background
ionosphere of the ray with given exit angle $\widehat{\Lambda}$.
The trajectory shape depends on the refractivity index distribution
$n(\overrightarrow{R}(\widehat{\Lambda},\sigma))$ in the ionosphere.
Traditionally, the calculations of rays for the decameter propagation
are made in hamiltonian optics approach \cite{Kravtsov_1990}. In
this approach, the ray trajectory is defined as a solution of the
Hamilton equations, and depends on the ray coordinates chosen. The
choice of ray coordinates depends on the convenience and on the structure
of Hamiltonian.

The eikonal $\psi$ in the equations (\ref{eq:GreensFunction},\ref{eq:RayRepresentationFormula},\ref{eq:KravcovMultipathFormula})
is also calculated from the Hamiltonian $H(\overrightarrow{P},\overrightarrow{R})$,
generalized momentum $\overrightarrow{P}$ and generalized coordinate
$\overrightarrow{R}$. One of the most used Hamiltonian $H(\overrightarrow{P},\overrightarrow{R})$
in geometrical optics is \cite{Kravtsov_1990}:

\begin{equation}
H(\overrightarrow{P},\overrightarrow{R})=\frac{1}{2}\left\{ P^{2}-n^{2}(\overrightarrow{R})\right\} =0\label{eq:ham_pn1}
\end{equation}

In this case the Hamilton equations, that allows to calculate the
ray trajectory and eikonal are:

\begin{equation}
\left\{ \begin{array}{c}
P=n(\overrightarrow{R})\\
\frac{\partial\overrightarrow{R}}{\partial S}=\overrightarrow{P}\\
\frac{\partial\overrightarrow{P}}{\partial S}=n\overrightarrow{\nabla}n(\overrightarrow{R})\\
\overrightarrow{\nabla}\psi=\overrightarrow{P}
\end{array}\right.\label{eq:gam_eq1}
\end{equation}

In this representation (\ref{eq:ham_pn1}) $S=\Lambda_{\Vert}$ is
ray coordinate, that is equivalent to the trajectory group length.
Eikonal, generalized coordinate and momentum are obtained by integrating
(\ref{eq:gam_eq1}) over the ray trajectory:

\begin{equation}
\left\{ \begin{array}{c}
\overrightarrow{R}=\overrightarrow{R}_{0}+\int\limits _{S_{0}}^{S}\overrightarrow{P}dS'\\
\overrightarrow{P}=\overrightarrow{P}_{0}+\int\limits _{S_{0}}^{S}n\overrightarrow{\nabla}n(\overrightarrow{R}(S'))dS'\\
\psi=\psi_{0}+\int\limits _{S_{0}}^{S}P^{2}dS'
\end{array}\right.\label{eq:gam_eq_solution}
\end{equation}

The hamiltonian optics representation is useful for numerical calculations
and is widely used in the problems of radiowave propagation (as characteristics
method)\cite{Kravtsov_1990,Settimi_2014}, including SuperDARN problems
\cite{Gauld_2002,Ponomarenko_2009,Ponomarenko_2010,Ponomarenko_2011,Berngardt_2015b}.

\subsection{Initial representation of the scattered signal}

One can see that ray representation of Green's function (\ref{eq:RayRepresentationFormula})
is the product of fast and slow oscillating functions. Spatial multiplier
$g(\widehat{\Lambda},\omega)$ is a smooth function of the frequency,
transmitted signal $a(\omega)$ is narrowband near carrier frequency
$\omega_{0}$. In this case we can represent fast oscillating phase
$\frac{\omega}{c}\psi(\overrightarrow{\Lambda},\omega)$ as a Taylor
series near $\omega_{0}$, with the small parameter $\left|\omega-\omega_{0}\right|\ll\omega_{0}$
and get the temporal representation of the propagating signal:

\begin{equation}
\begin{array}{c}
u(\overrightarrow{r},t)\approx e^{i\omega_{0}t}U(\overrightarrow{r},t)\\
U(\overrightarrow{r},t)\approx \int A(t-T_{r})g(\widehat{\Lambda},\omega_{0})F^{1/2}(\overrightarrow{\Lambda},\omega_{0})e^{ik_{0}\psi(\overrightarrow{\Lambda},\omega_{0})}\delta(\overrightarrow{r}-\overrightarrow{R}(\overrightarrow{\Lambda},\omega_{0}))d\overrightarrow{\Lambda}
\end{array}\label{eq:signal_propagation_narrowband}
\end{equation}

Here $k_{0}=\frac{\omega_{0}}{c}$ is wavenumber of transmitted signal;$U(\overrightarrow{r},t)$
is the complex envelope of propagating signal; $A(t-T_{r})$ is complex
envelope of the sounding signal;

\begin{equation}
T_{r,t}=\frac{\partial}{\partial\omega}\left(\frac{\omega}{c}\psi(\overrightarrow{\Lambda},\omega)\right)_{\omega=\omega_{0}}\label{eq:eikonal_delay}
\end{equation}

is the group delay of the propagating signal.

Necessary condition of narrowband signal (with spectral width $\Delta\omega$)
is defined as smallness of second differential of fast oscillating
phase in temporal representation of (\ref{eq:RayRepresentationFormula}):

\begin{equation}
(\Delta\omega)^{2}\frac{\partial^{2}}{\partial\omega^{2}}\left(\frac{\omega}{c}\psi(\overrightarrow{\Lambda},\omega)\right)_{\omega=\omega_{0}}\ll1\label{eq:narrowband_condition}
\end{equation}

So, the obtained equation (\ref{eq:signal_propagation_narrowband})
is valid in the regions, where antenna pattern, refractivity index
and geometrooptical divergence are changing slowly over the band $\Delta\omega$
of the sounding signal $a(\omega)$.

\begin{figure}
\includegraphics[scale=0.25]{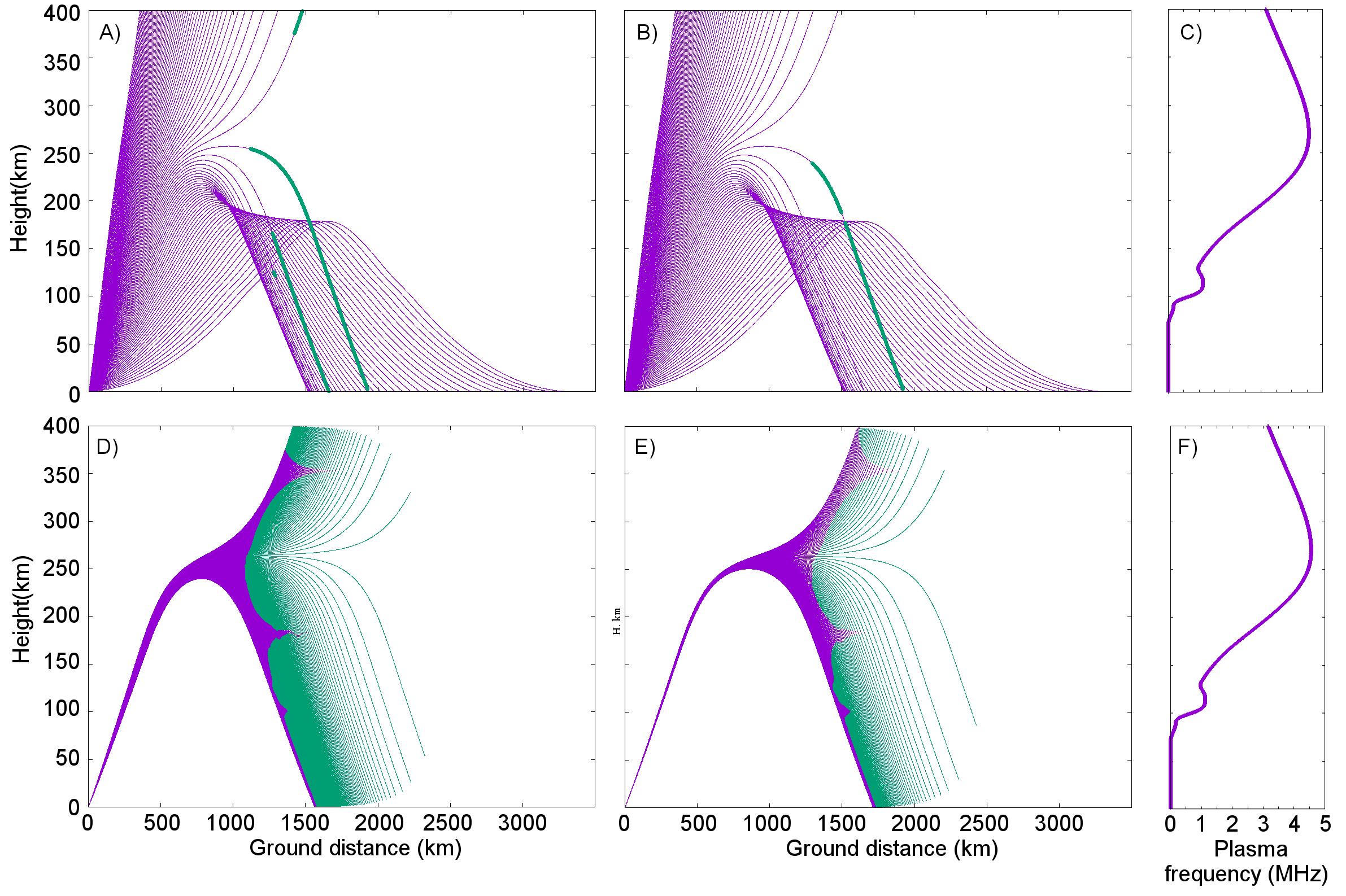}

\protect\protect\protect\caption{Regions of narrowband approximation validity (A,B,D,E) and ionospheric
profile (C,F) used for modeling. Blue color corresponds to the validity
region for initial equations (\ref{eq:signal_propagation_narrowband}).
Pulse duration is A,D) - 100 usec; B,E) - 300usec. A,B) - wide exit
ray angles range, D,E) - only trajectories near Pedersen ray.}

\label{fig:1} 
\end{figure}

In Fig.\ref{fig:1} are shown the validity (\ref{eq:narrowband_condition})
regions calculated for reference ionosphere model, typical SuperDARN
frequency (10MHz) and typical radar pulse durations (100 and 300usec.).
The signal trajectories were calculated by the method of characteristics
for cold isotropic plasma given by IRI-2012 model \cite{Bilitza_2011}.
As one can see from Fig.\ref{fig:1}, the approximation of the narrowband
signal (\ref{eq:narrowband_condition}) is valid almost everywhere,
except for a small number of rays near the Pedersen ray after reflection
point. For longer pulses (300usec), this area is smaller than for
shorter ones (100usec). Qualitatively, this can be explained as follows:
near the Pedersen ray small changes of signal frequency will lead
to a big changes of the trajectory (high frequency components of the
signal propagates upward, and the low frequency components propagates
downward). Thus, when propagating over these trajectories the signal
envelope in the ionosphere is not the same as the transmitted signal
envelope, and one should whether do not consider any of these areas,
or take into account the signal distortions at these trajectories
correctly.

Below in the paper we consider only the trajectories at which the
narrowband condition (\ref{eq:narrowband_condition}) is valid, and
use in (\ref{eq:signal_propagation_narrowband}) the equality sign.

By the same way we can obtain the expression for the scattered signal
at the receiving antenna:

\begin{equation}
\begin{array}{c}
U(t)\approx k_{0}^{2}\int\varepsilon(\overrightarrow{r},t-T_{t})A(t-T_{r}-T_{t})D(\overrightarrow{\Lambda}_{r},\overrightarrow{\Lambda}_{t},\omega_{0})e^{ik_{0}\Psi(\overrightarrow{\Lambda}_{r},\overrightarrow{\Lambda}_{t},\omega_{0})}\cdot\\
\cdot\delta(\overrightarrow{r}-\overrightarrow{R}_{r}(\overrightarrow{\Lambda}_{r},\omega_{0}))\delta(\overrightarrow{r}-\overrightarrow{R}_{t}(\overrightarrow{\Lambda}_{t},\omega_{0}))d\overrightarrow{\Lambda}_{t}d\overrightarrow{\Lambda}_{r}d\overrightarrow{r}
\end{array}\label{eq:signal_propagation_narrowband-1}
\end{equation}

where

$\overrightarrow{\Lambda}_{r},\overrightarrow{\Lambda}_{t}$ - ray
coordinate systems relative to receiver and transmitter correspondingly;

$\psi_{r}(\overrightarrow{\Lambda}_{r},\omega_{0}),\psi_{t}(\overrightarrow{\Lambda}_{t},\omega_{0})$
are the trajectory phase lengths from the scatterer position $\overrightarrow{R}_{r}(\overrightarrow{\Lambda}_{r},\omega_{0})$
to the receiver and transmitter correspondingly;

$\Psi(\overrightarrow{\Lambda}_{r},\overrightarrow{\Lambda}_{t},\omega_{0})=\psi_{r}(\overrightarrow{\Lambda}_{r},\omega_{0})+\psi_{t}(\overrightarrow{\Lambda}_{t},\omega_{0})$
is the full trajectory phase length (eikonal);

\begin{equation}
D(\overrightarrow{\Lambda}_{r},\overrightarrow{\Lambda}_{t},\omega_{0})=g_{r}^{*}(\widehat{\Lambda}_{r},\omega_{0})g_{t}(\widehat{\Lambda}_{t},\omega_{0})F_{r}^{1/2}(\overrightarrow{\Lambda}_{r},\omega_{0})F_{t}^{1/2}(\overrightarrow{\Lambda}_{t},\omega_{0})\label{eq:D_def}
\end{equation}
is the spatial multiplier, defined by antenna patterns $g_{r}(\widehat{\Lambda}_{r},\omega_{0}),g_{t}(\widehat{\Lambda}_{t},\omega_{0})$
of receiver and transmitter correspondingly, and by geometrical divergence;
indexes $t,r$ mark trajectories of transmitting (trajectory from
transmitter to the scatterer) and receiving (trajectory from receiver
to scatterer) correspondingly; $\overrightarrow{R}_{t}(\overrightarrow{\Lambda_{t}}),\overrightarrow{R}_{r}(\overrightarrow{\Lambda_{r}})$
are the positions of the scatterer as a function of ray coordinates
(ray direction and radar range) in transmitter coordinate system and
receiver coordinate system correspondingly; $T_{r},T_{t}$ are group
delays (\ref{eq:eikonal_delay}) from scatterer to transmitter and
receiver correspondingly; delta-functions define the condition that
scatterer position corresponds to the coincidence of the ends of trajectories
from transmitter and receiver (and at the same time are keeping the
ability of several propagation trajectories to the scatterer); $\varepsilon(\overrightarrow{r},t)$
is dielectric permittivity variations.

Integrating over the $d\overrightarrow{r}$ allows to simplify the
expression (\ref{eq:signal_propagation_narrowband-1}) and represent
it as an integral over the ray coordinates of transmitting and receiving
rays:

\begin{equation}
\begin{array}{c}
U(t)=k_{0}^{2}\int\varepsilon(\overrightarrow{R}_{t}(\overrightarrow{\Lambda}_{t},\omega_{0}),t-T_{t})A(t-T_{r}-T_{t})D(\overrightarrow{\Lambda}_{r},\overrightarrow{\Lambda}_{t},\omega_{0})\cdot\\
\cdot e^{ik_{0}\Psi(\overrightarrow{\Lambda}_{r},\overrightarrow{\Lambda}_{t},\omega_{0})}\delta(\overrightarrow{R}_{r}(\overrightarrow{\Lambda}_{r},\omega_{0})-\overrightarrow{R}_{t}(\overrightarrow{\Lambda}_{t},\omega_{0}))d\overrightarrow{\Lambda}_{t}d\overrightarrow{\Lambda}_{r}
\end{array}\label{eq:signal_scattering_narrowband}
\end{equation}

Delta-function under the integral in the case of several propagation
trajectories to the scatterer will transformed into the cross-sum
over these trajectories after integrating over the $\overrightarrow{\Lambda}_{t},\overrightarrow{\Lambda}_{r}$.

It is important to note that this equation is valid for narrowband
sounding and received signals (\ref{eq:narrowband_condition}) (where
$\Delta\omega$ is the sum of the bands of sounding signal $a(\omega)$
and dielectric permittivity fluctuations $\varepsilon(\omega)$) and
in absence of caustics for given group delay $t$.

For obtaining SuperDARN radar equations we will consider this equation
in the cases of monostatic and bistatic sounding.

\section{Monostatic sounding: single trajectory case}

In monostatic case the position of the receiver and the transmitter
coincides, so the functions $\overrightarrow{R}_{t}(\overrightarrow{\Lambda_{t}}),\overrightarrow{R}_{r}(\overrightarrow{\Lambda_{r}})$
coincides too and can be represented as a same function $\overrightarrow{R}(\overrightarrow{\Lambda})$,
but from different ray arguments $\overrightarrow{\Lambda_{t}},\overrightarrow{\Lambda_{r}}$
. In this case the expression for the received signal (\ref{eq:signal_scattering_narrowband})
becomes:

\begin{equation}
\begin{array}{c}
U(t)=k_{0}^{2}\int\varepsilon(\overrightarrow{R}(\overrightarrow{\Lambda}_{t}),t-T_{t})A(t-T_{r}-T_{t})D(\overrightarrow{\Lambda}_{r},\overrightarrow{\Lambda}_{t},\omega_{0})\cdot\\
\cdot e^{ik_{0}\Psi(\overrightarrow{\Lambda}_{r},\overrightarrow{\Lambda}_{t})}\delta(\overrightarrow{R}(\overrightarrow{\Lambda}_{r})-\overrightarrow{R}(\overrightarrow{\Lambda}_{t}))d\overrightarrow{\Lambda}_{t}d\overrightarrow{\Lambda}_{r}
\end{array}\label{eq:signal_scattering_singlepos}
\end{equation}

Here we excluded all the evident arguments of the functions. The expression
(\ref{eq:signal_scattering_singlepos}) describes both single and
multiple trajectory cases for the case of monostatic sounding.

\subsection{Single trajectory case}

To illustrate the approach of obtaining radar equation lets analyze
the simplest case: there is only a single trajectory to reach scatterer
from the transmitter. This corresponds to the case when $\overrightarrow{R}(\overrightarrow{\Lambda}_{r})$
is bijective function. In this case the integral over the $d\overrightarrow{\Lambda}_{r}$
in (\ref{eq:signal_scattering_singlepos}) can be calculated. The
rest argument $\overrightarrow{\Lambda}_{t}$ can be renamed to $\overrightarrow{\Lambda}$,
additional multiplier $F_{r}^{-1}$ arises due to integration of delta-function,
and the expression for the scattered signal (\ref{eq:signal_scattering_singlepos})
becomes:

\begin{equation}
U(t)=\int\varepsilon(\overrightarrow{R}(\overrightarrow{\Lambda}),t-T(\overrightarrow{\Lambda}))A(t-2T(\overrightarrow{\Lambda}))D_{u}(\overrightarrow{\Lambda},\omega_{0})e^{i2k_{0}\Psi(\overrightarrow{\Lambda})}d\overrightarrow{\Lambda}\label{eq:signal4-1}
\end{equation}

where

\begin{equation}
D_{u}(\overrightarrow{\Lambda}_{r},\omega_{0})=g_{r}^{*}(\widehat{\Lambda},\omega_{0})g_{t}(\widehat{\Lambda},\omega_{0})\label{eq:D_def-1}
\end{equation}

Usually, when interpreting SuperDARN data, the average autocorrelation
function of the signal\cite{Greenwald_1995,Chisham_2007,Ribeiro_2013}
is analysed:

\begin{equation}
P(t,\Delta T)=\left\langle U(t)U^{*}(t+\Delta T)\right\rangle \label{eq:P_UU}
\end{equation}

To get the radar equation, by analogy with \cite{Berngardt_1999,Berngardt_2000},
lets use the following characteristics:

\begin{equation}
\Phi(\overrightarrow{r};\overrightarrow{\rho},\Delta T)=\left\langle \varepsilon(\overrightarrow{r},t)\varepsilon^{*}(\overrightarrow{r}+\overrightarrow{\rho},t+\Delta T)\right\rangle \label{eq:irreg-spec-dens-def}
\end{equation}

is the stationary spatio-temporal correlation function of the irregularities,
that defines statistical characteristics of the scatterers;

\begin{equation}
W\left(t,S,\Delta T,\Delta S\right)=A(t-S/c)A^{*}(t-S/c+\Delta T+\Delta S/c)\label{eq:weight-volume-def}
\end{equation}

is the weight volume, that defines spatial selection of the irregularities
over the radar range (defined by two-way distance $S$) and selection
of the irregularities over the correlation function lag (defined by
parameter $\Delta T$ ) ;

\begin{equation}
\begin{array}{l}
D_{\Sigma}(\overrightarrow{\Lambda},\overrightarrow{\Lambda}_{2})=D_{u}(\overrightarrow{\Lambda},\omega_{0})D_{u}^{*}(\overrightarrow{\Lambda}_{2},\omega_{0})=\\
=g_{r}^{*}(\widehat{\Lambda}_{r},\omega_{0})g_{t}(\widehat{\Lambda}_{t},\omega_{0})g_{r}^{*}(\widehat{\Lambda}_{r,2},\omega_{0})g_{t}(\widehat{\Lambda}_{t,2},\omega_{0})\\
\\
\end{array}\label{eq:d-sum-def}
\end{equation}

is spatial multiplier that defines the contribution of antenna patterns.

The average autocorrelation function of the received signal (\ref{eq:P_UU})
can be represented as following temporal and spectral radar equations:

\begin{equation}
\begin{array}{c}
P(t,\Delta T)=\int\Phi(\overrightarrow{R}(\overrightarrow{\Lambda});\overrightarrow{R}(\overrightarrow{\Lambda}')-\overrightarrow{R}(\overrightarrow{\Lambda}),\Delta T-T(\overrightarrow{\Lambda}')+T(\overrightarrow{\Lambda}))\cdot\\
\cdot W(t,T(\overrightarrow{\Lambda})c,\Delta T,\left\{ T(\overrightarrow{\Lambda}')-T(\overrightarrow{\Lambda})\right\} c)D_{\Sigma}(\overrightarrow{\Lambda},\overrightarrow{\Lambda}')e^{i2k_{0}\left\{ \psi(\overrightarrow{\Lambda})-\psi(\overrightarrow{\Lambda}')\right\} }d\overrightarrow{\Lambda}d\overrightarrow{\Lambda}'
\end{array}\label{eq:signal4-1-2-1-1}
\end{equation}

\begin{equation}
\begin{array}{c}
\widetilde{P}(t,\omega)=\int\widetilde{\widetilde{\Phi}}\left(\overrightarrow{R}(\overrightarrow{\Lambda});\overrightarrow{k},\nu\right)\widetilde{W}\left(t,T(\overrightarrow{\Lambda})c,\omega-\nu,\left\{ T(\overrightarrow{\Lambda}')-T(\overrightarrow{\Lambda})\right\} c\right)\cdot\\
\cdot e^{i\nu\left\{ T(\overrightarrow{\Lambda})-T(\overrightarrow{\Lambda}')\right\} }D_{\Sigma}(\overrightarrow{\Lambda},\overrightarrow{\Lambda}')e^{-i\overrightarrow{k}\left(\overrightarrow{R}(\overrightarrow{\Lambda})-\overrightarrow{R}(\overrightarrow{\Lambda}')\right)}e^{i2k_{0}\left(\psi(\overrightarrow{\Lambda})-\psi(\overrightarrow{\Lambda}')\right)}d\overrightarrow{\Lambda}d\overrightarrow{\Lambda}'d\nu d\overrightarrow{k}
\end{array}\label{eq:ResSpectrum1}
\end{equation}

Here $\widetilde{A}$ marks Fourier transform of $A$ over one of
its arguments; $\widetilde{\widetilde{\Phi}}\left(\overrightarrow{R}(\overrightarrow{\Lambda});\overrightarrow{k},\nu\right)$
is spectral density of the irregularities.

It is difficult to use these equations for interpreting the experimental
data because of the fast oscillating functions under integral. But,
some of the integrands in (\ref{eq:signal4-1-2-1-1},\ref{eq:ResSpectrum1})
do not depend on $\overrightarrow{\Lambda}'$, the other can be considered
slightly dependent on it. So we can successfully integrate the equations:
at first over the $d\overrightarrow{\Lambda}'$ , and than over the
$d\overrightarrow{\Lambda}$.

To obtain a convenient representation of (\ref{eq:signal4-1-2-1-1},\ref{eq:ResSpectrum1})
we need to integrate fast oscillating functions like:

\begin{equation}
I=\int Z(\overrightarrow{\Lambda})e^{i\left(-\overrightarrow{k}\overrightarrow{R}(\overrightarrow{\Lambda},\omega_{0})+2k_{0}\psi(\overrightarrow{\Lambda},\omega_{0})\right)}d\overrightarrow{\Lambda}\label{eq:integral2}
\end{equation}

where $Z(\overrightarrow{\Lambda})$ is a function that is smooth
in comparison with fast-oscillating exponent. By analogy with \cite{Berngardt_1999,Berngardt_2000}
lets use a combination of the stationary phase method \cite{Fedoryuk_1989}
and Fourier transform for this.

\subsection{Integrating fast oscillating functions}

\subsubsection{Convenient ray coordinates}

The initial equations for the scattered signal (\ref{eq:ResSpectrum1})
are given in their basic representation, without specifying the exact
ray coordinate system $\overrightarrow{\Lambda}$. Different ones
can be used for this \cite{Kravtsov_1990}. The fast oscillating
integrand (\ref{eq:integral2}) in hamiltonian representation for
case (\ref{eq:gam_eq_solution}) becomes:

\begin{equation}
I=\int Z(\overrightarrow{\Lambda})e^{i\left(\int\limits _{0}^{S}\left\{ -\overrightarrow{k}\overrightarrow{P}(\hat{\Lambda},S')+2k_{0}P^{2}(\hat{\Lambda},S')\right\} dS'\right)}d\widehat{\Lambda}_{\xi}d\widehat{\Lambda}_{\eta}dS\label{eq:integral4}
\end{equation}

The most convenient technique for the calculation of fast oscillating
integrals is the method of stationary phase (MSP)\cite{Fedoryuk_1989}.
In this case, the integral can be expressed as the sum of the contributions
of the stationary points (points at which the derivative of fast-changing
phase over the integration variable becomes zero), and the contribution
is inversely proportional to the square root of second derivative
of the fast oscillating phase at this point (characteristic scale).
The criterion for the applicability of this method is small changes
of the other integrands at the characteristic scale \cite{Fedoryuk_1989}.

To check convenience of the ray coordinate system lets calculate the
integral (\ref{eq:integral4}) by stationary phase method over the
ray exit angles $\hat{\Lambda}=(\widehat{\Lambda}_{\xi},\widehat{\Lambda}_{\eta})$
that is convenient for this problem \cite{Berngardt_1999,Berngardt_2000}.

The detailed calculation by stationary phase method showed, that it
is convenient to choose the coordinates system that provides

\begin{equation}
\left\{ \overrightarrow{P}\frac{\partial\overrightarrow{P}}{\partial\hat{\Lambda}}\right\} _{\Lambda_{\Vert}=const}=0\label{eq:P_dPdL}
\end{equation}

that is valid for refraction-free or homogeneous background ionosphere.
In this case obtaining and interpreting the radar equations will be
similar to the approach developed for VHF \cite{Berngardt_1999}.
Due to the generalized momentum $\overrightarrow{P}$ is perpendicular
to the wave front at any given moment \cite[p.24]{Babich_1972}, we
should use instead of the trajectory length coordinate $\Lambda_{\parallel}$
the trajectory group length $S$ (or group delay of the signal $dT=\frac{dL}{cn}$),
that provides (\ref{eq:P_dPdL}). In this case, one can use the system
of Hamiltonian equations, described above (\ref{eq:gam_eq1},\ref{eq:gam_eq_solution})
and analyze the integral (\ref{eq:integral4}).

Using (\ref{eq:gam_eq1},\ref{eq:gam_eq_solution}) it can be shown
that stationary point condition $\frac{\partial\varphi}{\partial\hat{\Lambda}}=0$
corresponds to:

\begin{equation}
\left\{ \frac{\partial}{\partial\hat{\Lambda}}\int\limits _{0}^{S}\left(\overrightarrow{k}\overrightarrow{P}\right)dS'\right\} _{\hat{\Lambda}=\hat{\Lambda}_{SF}}=\int\limits _{0}^{S}\left\{ \overrightarrow{k}\left(\frac{\partial\overrightarrow{P}}{\partial\hat{\Lambda}}\right)_{\bot\overrightarrow{P}}\right\} _{\hat{\Lambda}=\hat{\Lambda}_{SF}}dS'=0\label{eq:StatConditionScalar}
\end{equation}

This means that stationary point is the point where the projection
of the irregularities wave vector $\overrightarrow{k}$ to the perpendicular
to the generalized momentum $\overrightarrow{P}$ 'in average' (as
an integral along the trajectory) equals to zero. In other words,
the irregularities wave vector $\overrightarrow{k}$ is 'in average'
should be perpendicular to the wave front.

So the stationary point (\ref{eq:StatConditionScalar}) for coordinate
system $(\hat{\Lambda},S)$ can be represented as:

\begin{equation}
\int\limits _{0}^{S}\overrightarrow{k}\overrightarrow{P}dS'=-\int\limits _{0}^{S}kPdS'\label{eq:ST_integral}
\end{equation}

For locally-homogeneous trajectory, i.e. for trajectory, at which
the refraction coefficient is nearly constant at investigated part
of the trajectory ($P=const$), the stationary point can be represented
not as the 'average' integral condition (\ref{eq:StatConditionScalar}),
but as the local condition:

\begin{equation}
\overrightarrow{k}\overrightarrow{P}=-kP\label{eq:ST_loc}
\end{equation}

This condition is structurally coincides with the refraction-free
case, analyzed in \cite{Berngardt_1999}, and corresponds the the
well-known Wolf-Bragg condition.

\subsubsection{Integrating over the ray angles}

By integrating (\ref{eq:ResSpectrum1}) over the ray exit angles $\hat{\Lambda},\hat{\Lambda}'$
by MSP and using stationary phase condition in form (\ref{eq:ST_integral}),
we obtain:

\begin{equation}
\begin{array}{c}
P(t,\Delta T)=\int\Phi\left(\overrightarrow{R}(\overrightarrow{\Lambda});\overrightarrow{k},\Delta T+S/c-S'/c\right)W_{2,t}\left(\hat{\Lambda}_{SF},S,\Delta T,S-S'\right)\cdot\\
\cdot\left|\gamma_{2}\right|^{2}e^{-i\int\limits _{S'}^{S}P\left(k-2k_{0}P(\hat{\Lambda}_{SF},S_{1},\omega_{0})\right)dS_{1}}dSdS'd\overrightarrow{k}
\end{array}\label{eq:autocorfunction1-2-2-1-1}
\end{equation}

where

\begin{equation}
\begin{array}{c}
W_{2,t}\left(\hat{\Lambda}_{SF},S,\Delta T,S-S'\right)=W\left(t,2S,\Delta T,2\left\{ S-S'\right\} \right)V_{\Lambda}^{-1}(\hat{\Lambda}_{SF},S,\omega_{0})\cdot\\
\cdot V_{\Lambda}^{-1}(\hat{\Lambda}_{SF},S',\omega_{0})D_{\Sigma}(\hat{\Lambda}_{SF},S)
\end{array}\label{eq:W_2t}
\end{equation}

is weight volume that defines spatio-temporal resolution of the technique;
$V_{\Lambda}$ is effective scattering volume; $\hat{\Lambda}_{SF}$
is stationary point over $\hat{\Lambda}$;

\[
\gamma_{2}=\left(2\pi\right)^{n/2}e^{i\frac{\pi}{4}sgn(S'')}=2\pi i
\]
is additional multiplier from stationary phase method \cite{Fedoryuk_1989}.

From the initial equation (\ref{eq:ResSpectrum1}) it is obvious that
for slowly changing $\Phi\left(\overrightarrow{R}(\hat{\Lambda}_{SF},S);\right)W\left(t-2T(\hat{\Lambda}_{SF},S)\right)$
the stationary points over $\hat{\Lambda},\hat{\Lambda}'$ are similar
and marked as $\hat{\Lambda}_{SF}$ , and satisfy the Wolf-Bragg condition
(\ref{eq:StatConditionScalar}).

The basic element of the calculations by stationary phase method is
the second differential of fast oscillating integrand function $\psi$.
It defines the integral contribution of the stationary point to the
integral, and inverse angular size of the region, making this contribution
\cite{Fedoryuk_1989}:

\begin{equation}
V_{\Lambda}=k\left\Vert \frac{\partial^{2}\psi}{\partial\hat{\Lambda}_{i}\partial\hat{\Lambda}_{j}}\right\Vert ^{1/2}\label{eq:WR-def}
\end{equation}

In geometrical optics the Hessian $V_{\Lambda}$ can be defined through
the product of two main curvature radii $R_{GM},R_{Gm}$ \cite{Kravtsov_1990}:

\begin{equation}
V_{\Lambda}=kF^{1/2}=k\left\{ R_{GM}R_{Gm}\right\} ^{1/2}\label{eq:WR_RgRg}
\end{equation}

In spectral representation the radar equation (\ref{eq:autocorfunction1-2-2-1-1})
becomes:

\begin{equation}
\begin{array}{c}
\widetilde{P}(t,\omega)=\int\widetilde{\widetilde{\Phi}}\left(\overrightarrow{R}(\overrightarrow{\Lambda});\overrightarrow{k},\nu\right)\widetilde{W}_{2,t}\left(\hat{\Lambda}_{SF},S,\omega-\nu,S-S'\right)\cdot\\
\cdot\left|\gamma_{2}\right|^{2}e^{i\nu\left\{ S/c-S'/c\right\} }e^{-i\int\limits _{S'}^{S}P\left(k-2k_{0}P(\hat{\Lambda}_{SF},S_{1},\omega_{0})\right)dS_{1}}dSdS'd\overrightarrow{k}d\nu
\end{array}\label{eq:RLU-Final-Integral}
\end{equation}

The criterion of applicability of the MSP is a small changes of integrands
$\widetilde{\widetilde{\Phi}}\left(\overrightarrow{R}(\overrightarrow{\Lambda});\overrightarrow{k},\nu\right),\widetilde{W}_{2,t}\left(\hat{\Lambda}_{SF},S,\omega-\nu,S-S'\right)$
at angles about $V_{\Lambda}^{-1}$ (\ref{eq:WR_RgRg}), or roughly
equivalent, small changes of the integrands at distances of the order
Fresnel zone $\sqrt{\lambda R}$. This condition has equivalent in
the refraction-free case \cite{Berngardt_1999}. It should be noted
that the smoothness of the refraction index on the characteristic
scale of the Fresnel zone is one of the conditions of applicability
of geometrical optics \cite{Kravtsov_1990}. Thus, the resulting
representation of radar equation (\ref{eq:RLU-Final-Integral}) is
valid almost everywhere, where the original equation (\ref{eq:signal4-1})
is valid and, at the same time, where the average spectral density
$\widetilde{\widetilde{\Phi}}\left(\overrightarrow{R}(\overrightarrow{\Lambda});\overrightarrow{k},\nu\right)$
varies smoothly over the spatial variable $\overrightarrow{R}(\overrightarrow{\Lambda})$.

\subsubsection{Integrating the radar equation over the range}

Lets calculate the integral from function (\ref{eq:RLU-Final-Integral})
that oscillates fast over the $\Delta S=S-S'$ with the phase $\varphi$:

\begin{equation}
\varphi=\nu\Delta S-\int\limits _{S-\Delta S}^{S}P\left(k-2k_{0}P(\hat{\Lambda}_{SF},S_{1},\omega_{0})\right)dS_{1}\label{eq:PhaseForRegularityCondition}
\end{equation}

As it will be shown, it is convenient to analyze the two extreme cases:
locally-homogeneous and locally-inhomogeneous one. The integration
domain over $\Delta S$ is limited by the weight volume $W_{2,t}\left(\hat{\Lambda}_{SF},S,\Delta T,\Delta S\right)$
carrier. For pulse sequences used at SuperDARN radars\cite{Chisham_2007}
this region is defined by elementary sounding pulse duration.

It can be shown that for SuperDARN radars the carrier of $W_{2,t}$
has a specific shape over $\Delta S$ - a peak near zero and a few
peaks at very large distances $\Delta S$. Generally, this volume
can be decomposed into the sum of a peaks. For large distances $\Delta S$
we integrate regions separated by thousands kilometers so the correlation
at large $\Delta S$ and integrals with large $\Delta S$ can be neglected.

Therefore lets consider the following two limiting cases: when it
is a linear function of $\Delta S$ (it is equivalent to the smallness
of second differential of the phase (\ref{eq:PhaseForRegularityCondition})
over $\Delta S$) and the integral can be calculated as Fourier transform,
and when (\ref{eq:PhaseForRegularityCondition}) is very nonlinear
function of $\Delta S$ (when the second differential of the phase
(\ref{eq:PhaseForRegularityCondition}) over $\Delta S$ is big enough)
and the integral can be calculated with the stationary phase method.

So the (\ref{eq:RLU-Final-Integral}) calculations can be differed
into two classes:

- the locally-homogeneous parts of trajectories (with nearly constant
refraction coefficient along the trajectory), at which the scattered
signal is accumulated coherently over $\Delta S$ near zero;

- the locally-inhomogeneous parts of trajectories, at which the main
contribution into the integral is made by relatively small area inside
the integration region $\Delta S$.

So the radar equation (\ref{eq:RLU-Final-Integral}) will be a sum
of two terms - the contributions from locally-homogeneous and from
locally-inhomogeneous trajectories.

\paragraph{Locally-homogeneous trajectory case}

In the case of locally-homogeneous trajectories the resulting signal
is defined by 2D stationary phase method over the ray exit directions
$\hat{\Lambda}$, and by the Fourier-transform over the radar range
$\Delta S$. The locally-homogeneous case is equivalent to the refraction-free
case \cite{Berngardt_1999}, and 2D stationary phase condition (\ref{eq:StatConditionScalar})
can be considered locally too (\ref{eq:ST_loc}).

It is obvious that background ionosphere can be considered as locally-homogeneous
for given sounding pulse only in the case of sufficiently short pulse.
However, it should satisfy the narrowband sounding requirement (\ref{eq:narrowband_condition}),
to initial formulas to be valid. When pulse becomes longer and longer
there comes a moment when the locally-homogeneous trajectory condition
stops working. In this case coherent accumulation of the signal stops
and scattered signal amplitude growing stops too. Qualitatively one
can suppose that maximal scattered power is reached when the pulse
duration and local-homogenety scale length becomes equal.

The locally-homogeneous trajectory condition (i.e. the maximal size
of the region, at which the phase $\varphi$ (\ref{eq:PhaseForRegularityCondition})
can be considered as linear function of $\Delta S$) in the first
approximation can be defined from smallness of the second differential
of phase (\ref{eq:PhaseForRegularityCondition}) as

\begin{equation}
\left(\frac{\Delta R_{A}}{J_{1}}\right)^{2},\left(\frac{\Delta T_{A}}{cJ_{1}}\right)^{2}<1\label{eq:LocallyRegularCondition}
\end{equation}

where $\Delta T_{A}$ is the elementary sounding pulse duration (defined
from the spatial resolution of sounding pulse $\Delta R_{A}$); $J_{1}$
- the characteristic scale of quadratic phase changes:

\begin{equation}
J_{1}=(2k_{0}P\frac{\partial P}{\partial S})^{-1/2}\label{eq:J1}
\end{equation}

If $\Delta L$ is a characteristic scale of changes of refractivity
index along the trajectory, then $J_{1}\sim\sqrt{\lambda_{0}\Delta L}$.

From the condition of locally-homogeneous trajectory (\ref{eq:LocallyRegularCondition})
one can see that the requirement is equivalent to small variations
of the background refraction coefficient along the trajectory and
to short sounding pulses. At these trajectories the radar equation
(\ref{eq:RLU-Final-Integral}) can be significantly simplified. The
integration over the $d(S-S')$ produces Fourier transform of $W_{2,t}$:

\begin{equation}
\begin{array}{c}
\widetilde{P}(t,\omega)=\int\widetilde{\widetilde{\Phi}}\left(\overrightarrow{R}(\overrightarrow{\Lambda});\overrightarrow{k},\nu\right)\cdot\\
\cdot\left|\gamma_{2}\right|^{2}\widetilde{\widetilde{W}}_{2,t}\left(\hat{\Lambda}_{SF},S,\omega-\nu,\left(kP(\hat{\Lambda}_{SF},S,\omega_{0})-2k_{0}P^{2}(\hat{\Lambda}_{SF},S,\omega_{0})\right)-\nu/c\right)dSd\overrightarrow{k}d\nu
\end{array}\label{eq:ResultRLU-quasihomogeneous}
\end{equation}

The narrowband of the sounding signal and spectral density of the
irregularities in this case determines the smallness of the wavenumbers
region involved into the scattering:

\begin{equation}
\left|kP(\hat{\Lambda}_{SF},S,\omega_{0})-2k_{0}P^{2}(\hat{\Lambda}_{SF},S,\omega_{0})\right|<\Delta\omega/c\label{eq:StatCondition-Local-mod}
\end{equation}

\begin{figure}
\includegraphics{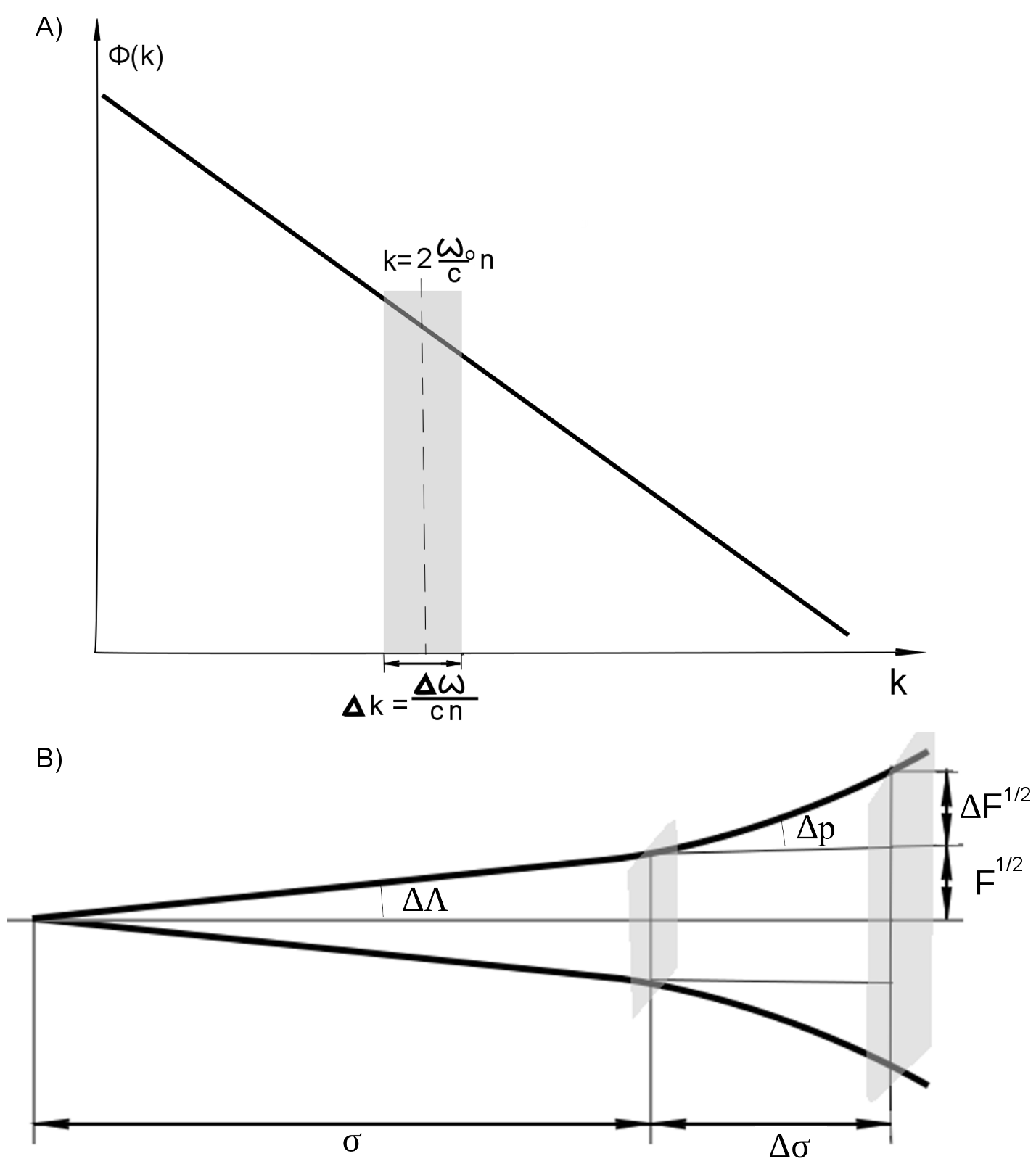} \protect\protect\protect\caption{A) Selective properties of scattering for the locally-homogeneous
monostatic case (\ref{eq:ResultRLU-quasihomogeneous},\ref{eq:ResultRLU-quasihomogeneous-dR}).
B) Geometrical considerations for calculating $J_{\hat{P}\rightarrow\hat{\Lambda}}$}

\label{fig:fig_dP} 
\end{figure}

Taken together, these two conditions (\ref{eq:StatCondition-Local-mod},\ref{eq:ST_loc})
define space-phase matching for radiowaves and irregularities (Wolf-Bragg
condition): the main contribution to the scattering in a smoothly
irregular media is produced by irregularities spatial harmonics with
wavevectors perpendicular to the wave front of the propagating radio
wave, and with the wavenumber nearly equal to the doubled wave number
of radiowave propagating in the medium. It is necessary to note that
the closer $P$ (or $n$) to 0, the wider the integration region over
$dk$. The conditions (\ref{eq:StatCondition-Local-mod},\ref{eq:ST_loc})
are illustrated at Fig.\ref{fig:fig_dP}A.

To represent (\ref{eq:ResultRLU-quasihomogeneous}) as an integral
over the space $d\overrightarrow{r}$, one needs to calculate Jacobian
for the transformation $dSd\hat{k}\rightarrow d\overrightarrow{r}$.
To simplify the calculations, lets take into account Wolf-Bragg condition
($d\hat{k}=d\hat{P}$) and calculate the following two transformations:
$dSd\hat{P}\rightarrow dSd\hat{\Lambda}\rightarrow d\overrightarrow{r}$
.

Geometrical interpretation of $F$ is cross sectional area of the ray\cite{Kravtsov_1990}. From geometrical considerations (see Fig.\ref{fig:fig_dP}B) it is
obvious that

\begin{equation}
J_{\hat{P}\rightarrow\hat{\Lambda}}\approx\left\{ \frac{1}{2}\frac{\sigma}{F}\frac{\partial F}{\partial\sigma}\right\} ^{2}\label{eq:J_P_L}
\end{equation}

It is also clear that $J_{dSd\hat{\Lambda}\rightarrow d\overrightarrow{r}}=2n^{-1}\left|\frac{\partial\overrightarrow{r_{i}}}{\partial\overrightarrow{\Lambda_{j}}}\right|^{-1}=2n^{-1}F^{-1}$.
So

\begin{equation}
\begin{array}{c}
\widetilde{P}(t,\omega)=\int\widetilde{\widetilde{\Phi}}\left(\overrightarrow{r};\overrightarrow{k},\nu\right)\left|\gamma_{2}\right|^{2}k^{2}\left\{ \frac{2}{nF}J_{\hat{P}\rightarrow\hat{\Lambda}}\right\} \cdot\\
\cdot\widetilde{\widetilde{W}}_{2,t}\left(\hat{\Lambda}_{SF}(\overrightarrow{r}),S(\overrightarrow{r}),\omega-\nu,\left(kP(\hat{\Lambda}_{SF}(\overrightarrow{r}),S(\overrightarrow{r}),\omega_{0})-2k_{0}P^{2}(\hat{\Lambda}_{SF}(\overrightarrow{r}),S(\overrightarrow{r}),\omega_{0})\right)-\nu/c\right)d\overrightarrow{r}dkd\nu
\end{array}\label{eq:ResultRLU-quasihomogeneous-dR}
\end{equation}

where $W_{2,t}\left(\hat{\Lambda}_{SF},S,\Delta T,S-S'\right)$ is
defined by (\ref{eq:W_2t}).

The additional spatial multiplier $\left\{ \frac{2}{nF}J_{\hat{P}\rightarrow\hat{\Lambda}}\right\} $
is relatively smooth over the space (except the caustics regions,
where it becomes infinite). For refraction-free case it is equal to
$\frac{2}{r^{2}}$, and the radar equation (\ref{eq:ResultRLU-quasihomogeneous-dR})
becomes similar to the obtained in \cite{Berngardt_1999}.

In Fig.\ref{fig:2} are illustrated the locally-homogeneous areas
for model ionosphere. The figure shows that such basic areas are:
field caustics associated with reflection (170-270km); area corresponding
F2-peak (270km) and E-peak (120 km); and the region of linear propagation
out of the ionosphere (area 0-90km altitude).

\paragraph{Locally-inhomogeneous trajectory case}

With increase of the variations of the refractivity index along the
trajectory or/and increase of the sounding pulse duration the situation
changes gradually. In this case the phase (\ref{eq:PhaseForRegularityCondition})
can not be considered as a linear function of $\Delta S$, and Fourier
transform approximation can not be used, and one need to use other
approximate methods to calculate fast oscillating integral (\ref{eq:RLU-Final-Integral}).

In the case of locally-inhomogeneous trajectories integration of the
radar equation (\ref{eq:RLU-Final-Integral}) over $\Delta S$ can
be done by using stationary phase method. One also can not use the
local condition for stationary point (\ref{eq:ST_loc}) instead of
the integral one (\ref{eq:ST_integral}). This mean that the stationary
point will be defined from 3D structure of the integral.

Calculating the position of the stationary point $\hat{\Lambda}_{SF},S_{SF}$
over $S$ and $S'$ separately in (\ref{eq:RLU-Final-Integral}) gives
the following integral conditions for stationary phase:

\begin{equation}
\left\{ \begin{array}{c}
\int\limits _{S-\Delta S_{W}}^{S}\overrightarrow{k}\left(\frac{\partial\overrightarrow{P}}{\partial\Lambda}\right)_{\bot\overrightarrow{P}}dS_{1}=0\\
\overrightarrow{k}\overrightarrow{P}(\hat{\Lambda}_{SF},S_{SF})-2k_{0}P^{2}(\hat{\Lambda}_{SF},S_{SF})-\nu/c=0
\end{array}\right.\label{eq:StatConditionScalar-LoclaKriv}
\end{equation}

In the case when the stationary point $S_{SF}$ is a single one and
is within the carrier of $W(S)$ $[S-\Delta S_{W},S]$ for given $t$
it makes a big contribution into integral (\ref{eq:RLU-Final-Integral}).
In other cases the contribution is small. It should be noted that
multiplier outside the carrier of W(S) $\int\limits _{0}^{S-\Delta S_{W}}\overrightarrow{k}\left(\frac{\partial\overrightarrow{P}}{\partial\Lambda}\right)_{\bot}dS_{1}$
can be removed out the integral sign. So actually the MSP (\ref{eq:StatConditionScalar-LoclaKriv})
is calculated over the carrier of W(S).

In this case both stationary points (for $S$ and $S'$) coincide
and integral will lead to the $W(\Delta S=0)$, and the expression
(\ref{eq:RLU-Final-Integral}) becomes:

\begin{equation}
\begin{array}{c}
P(t,\Delta T)=\int\widetilde{\Phi}\left(\overrightarrow{R}(\overrightarrow{\Lambda}_{SF});\overrightarrow{k},\Delta T\right)W\left(t,2T(\overrightarrow{\Lambda}_{SF})c,\Delta T,0\right)\cdot\\
\cdot\left|\gamma_{2}\gamma_{1}\right|^{2}J_{1}^{2}V_{\Lambda}^{-2}(\overrightarrow{\Lambda}_{SF},\omega_{0})D_{\Sigma}(...)d\overrightarrow{k}
\end{array}\label{eq:autocorfunction1-2-2-2}
\end{equation}

\begin{equation}
\begin{array}{c}
\widetilde{P}(t,\omega)=\int\widetilde{\widetilde{\Phi}}\left(\overrightarrow{R}(\overrightarrow{\Lambda}_{SF});\overrightarrow{k},\nu\right)\widetilde{W}\left(t,2T(\overrightarrow{\Lambda}_{SF})c,\omega-\nu,0\right)\cdot\\
\cdot\left|\gamma_{2}\gamma_{1}\right|^{2}J_{1}^{2}V_{\Lambda}^{-2}(\overrightarrow{\Lambda}_{SF},\omega_{0})D_{\Sigma}(...)d\overrightarrow{k}d\nu
\end{array}\label{eq:autocorfunction1-2-2-2-1}
\end{equation}

where 
\[
\gamma_{1}=\left(2\pi\right)^{1/2}e^{i\frac{\pi}{4}}=\sqrt{2\pi i}
\]
is additional multiplier from stationary phase method \cite{Fedoryuk_1989};
$J_{1}$ is defined by (\ref{eq:J1}).

Taking into account the same considerations, as for analysis formula
(\ref{eq:ResultRLU-quasihomogeneous}), we can obtain:

\begin{equation}
\begin{array}{c}
P(t,\omega)=\int\widetilde{\widetilde{\Phi}}\left(\overrightarrow{R}(\hat{\Lambda}_{SF},S_{SF}(\hat{k} _{SF}k));\hat{k} _{SF} k,\nu\right)\widetilde{W}_{2,t}\left(\hat{\Lambda}_{SF},T(\hat{\Lambda}_{SF},S_{SF}(\hat{k} _{SF}k))c,\omega-\nu,0\right)\cdot\\
\cdot\left|\gamma_{2}\gamma_{1}\right|^{2}J_{1}^{2}(\hat{\Lambda}_{SF},S_{SF}(\hat{k} _{SF}k))D_{\Sigma}(...)J_{\hat{P}\rightarrow\hat{\Lambda}}k^{2}dkd\hat{\Lambda}d\nu
\end{array}\label{eq:RLU-LocInhom}
\end{equation}

To estimate the relation between contributions of locally-homogeneous
(\ref{eq:ResultRLU-quasihomogeneous}) and locally-inhomogeneous (\ref{eq:RLU-LocInhom}),
we can estimate $d\overrightarrow{r}\sim\left(\frac{S}{2}\right)^{2}\Delta R_{A}d\hat{\Lambda}$
in (\ref{eq:ResultRLU-quasihomogeneous}). So the relation can be
estimated as:

\[
\frac{P_{inhom}(t,\omega)}{P_{hom}(t,\omega)}\approx\frac{J_{1}^{2} J_{\hat{P}\rightarrow\hat{\Lambda}}\frac{\Delta\omega_{A}}{nc}\Delta\omega_{A}}{\left(\frac{S}{2}\right)^{2}\left\{ \frac{2}{nF}J_{\hat{P}\rightarrow\hat{\Lambda}}\right\} \Delta R_{A}^{2}\frac{\Delta\omega_{A}}{nc}\Delta\omega_{A}}\approx\frac{J_{1}^{2}}{\{\Delta R_{A}\}^{2}}
\]

So the energy of the scattered signal in locally-homogeneous case
approximately $\left(\Delta R_{A}/J_{1}\right)^{2}$ times higher
then in locally-inhomogeneous case.

In Fig.\ref{fig:3}A shown the value of $J_{1}^{2}/\left(\Delta R_{A}\right)^{2}$
for locally-homogeneous (=0dB, yellow areas), and for locally-inhomogeneous
trajectories (red and blue colors). The calculations for Fig.\ref{fig:3}A
are made with the reference model ionosphere (IRI-2012) and typical
frequency and duration of SuperDARN sounding signal.

In Fig.\ref{fig:3}A shown that the regions with locally-inhomogeneous
trajectories (red and blue) are located between the E- and F-maximum
and expected to produce scattering about 10-15dB less powerful than
the region of locally-homogeneous trajectories. Because of the field-aligned
scatterers (studied by SuperDARN radars) produce powerful aspect scattering
these areas look still important for future consideration.

\section{Bistatic sounding}

To analyze the bistatic sounding we assume, by analogy with \cite{Berngardt_2000}
that trajectories of the transmitted and received signal are spaced
far enough from each other. This means that the angle between the
incident and scattered ray bigger than the angular size of the area
of the stationary phase (\ref{eq:WR_RgRg}). The case of smaller angle
the problem degenerates into the monostatic case, already analyzed
(\ref{eq:autocorfunction1-2-2-2}). It should be noted that in presence
of refraction this condition in general is not connected with the
relative position of transmitter and receiver (in opposite to refraction-free
situation \cite{Berngardt_2000}), but is connected only with the
angle between the trajectories of transmitted and scattered signal
at the point of scattering.

\subsection{Single trajectory pair: initial equation}

Lets analyze the case of single trajectory pair: there is only one
propagation path from the transmitter to scattering point, and only
one path from receiver to the scattering point. And they differ.

We consider the scattered signal as the integral over the ray coordinates
$\overrightarrow{\Lambda}_{t}$ associated with the transmitter and
over the spatial coordinates $\overrightarrow{r}$ of the scatterer
position. In this case, the scattering signal (\ref{eq:signal_scattering_narrowband})
leads to the following correlation function $P(t,\Delta T)$:

\begin{equation}
\begin{array}{c}
P(t,\Delta T)=\int\widetilde{\Phi}(\overrightarrow{r};\overrightarrow{k},\Delta T+T_{t}-T_{t}')\cdot\\
\cdot W\left(t,\left\{ T_{r}+T_{t}\right\} c,\Delta T,\left\{ \left(T_{r}+T_{t}\right)-\left(T_{r}'+T_{t}'\right)\right\} c\right)D(\overrightarrow{\Lambda}_{r},\overrightarrow{\Lambda}_{t},\omega_{0})D(\overrightarrow{\Lambda}_{r}',\overrightarrow{\Lambda}_{t}',\omega_{0})\cdot\\
\cdot e^{i\overrightarrow{k}\left(\overrightarrow{r}-\overrightarrow{r}'\right)}e^{ik_{0}\Psi(\overrightarrow{\Lambda}_{r},\overrightarrow{\Lambda}_{t})}e^{-ik_{0}\Psi(\overrightarrow{\Lambda}_{r}',\overrightarrow{\Lambda}_{t}')}\cdot\\
\cdot\delta(\overrightarrow{r}-\overrightarrow{R}(\overrightarrow{\Lambda}_{t}))\delta(\overrightarrow{r}'-\overrightarrow{R}(\overrightarrow{\Lambda}_{t}'))F^{-1}(\overrightarrow{\Lambda}_{r}(\overrightarrow{r}))F^{-1}(\overrightarrow{\Lambda}_{r}'(\overrightarrow{r}'))d\overrightarrow{\Lambda}_{t}d\overrightarrow{r}d\overrightarrow{\Lambda}_{t}'d\overrightarrow{r}'d\overrightarrow{k}
\end{array}\label{eq:RLU-result-3D-3}
\end{equation}

\subsection{Single trajectory pair: obtaining of the radar equation}

Integrating delta-functions over the $\overrightarrow{\Lambda}_{t},\overrightarrow{\Lambda}_{t}'$
in (\ref{eq:RLU-result-3D-3}) will result to $\overrightarrow{\Lambda}_{t}(\overrightarrow{r})$
and $\overrightarrow{\Lambda}_{t}'(\overrightarrow{r}')$ functions
under the integral and corresponding Jacobian $F^{-1}$:

\begin{equation}
\begin{array}{c}
P(t,\Delta T)=\int\widetilde{\Phi}(\overrightarrow{r};\overrightarrow{k},\Delta T+T_{t}-T_{t}')\\
W\left(t,\left\{ T_{r}+T_{t}\right\} c,\Delta T,\left\{ \left(T_{r}+T_{t}\right)-\left(T_{r}'+T_{t}'\right)\right\} c\right)D(\overrightarrow{\Lambda}_{r},\overrightarrow{\Lambda}_{t},\omega_{0})D(\overrightarrow{\Lambda}_{r}',\overrightarrow{\Lambda}_{t}',\omega_{0})\\
e^{i\overrightarrow{k}\left(\overrightarrow{r}-\overrightarrow{r}'\right)}e^{ik_{0}\Psi(\overrightarrow{\Lambda}_{r},\overrightarrow{\Lambda}_{t})}e^{-ik_{0}\Psi(\overrightarrow{\Lambda}_{r}',\overrightarrow{\Lambda}_{t}')}\\
\left|\gamma_{3}\right|^{2}F_{4}^{-1}(\overrightarrow{r}(\overrightarrow{k}),\overrightarrow{r}'(\overrightarrow{k}))d\overrightarrow{r}d\overrightarrow{r}'d\overrightarrow{k}
\end{array}\label{eq:RLU-result-3D-3-1}
\end{equation}

where 
\begin{equation}
F_{4}(\overrightarrow{r},\overrightarrow{r}')=F(\overrightarrow{\Lambda}_{r}(\overrightarrow{r}))F(\overrightarrow{\Lambda}_{r}'(\overrightarrow{r}'))F(\overrightarrow{\Lambda}_{t}(\overrightarrow{r}))F(\overrightarrow{\Lambda}_{t}'(\overrightarrow{r}'))\label{eq:F4}
\end{equation}

\[
\gamma_{3}=\left(2\pi\right)^{n/2}e^{i\frac{\pi}{4}sgn(S'')}=\left(2\pi i\right)^{3/2}
\]
is additional multiplier from MSP \cite{Fedoryuk_1989}.

We can integrate the expression (\ref{eq:RLU-result-3D-3-1}) over
the $d\overrightarrow{r}$ and $d\overrightarrow{r}'$ by MSP separately.

Integrating over $d\overrightarrow{r}$ will lead to the following
stationary phase condition:

\begin{equation}
\overrightarrow{k}+k_{0}\frac{\partial}{\partial\overrightarrow{r}}\psi_{t}(\overrightarrow{\Lambda}_{t})+k_{0}\frac{\partial}{\partial\overrightarrow{r}}\psi_{r}(\overrightarrow{\Lambda}_{r})=0
\end{equation}

Taking into account that $\overrightarrow{r}=\overrightarrow{R_{r}}(\overrightarrow{\Lambda}_{r})=\overrightarrow{R_{t}}(\overrightarrow{\Lambda}_{t})$
and taking into account the properties of hamiltonian trajectories
(\ref{eq:gam_eq_solution}), the stationary phase condition will take
the clear look that shows the selective properties of the scattering:

\begin{equation}
\overrightarrow{k}+k_{0}\overrightarrow{P}_{t}(\overrightarrow{\Lambda}_{t})+k_{0}\overrightarrow{P}_{r}(\overrightarrow{\Lambda}_{r})=0\label{eq:WD-conditon-2st}
\end{equation}

Doing the same over $d\overrightarrow{r}'$and assuming the stationary
point $\overrightarrow{r}^{0}$ to be only one for each $\overrightarrow{k}$
we can transform the radar equation (\ref{eq:RLU-result-3D-3-1})
to the form:

\begin{equation}
\begin{array}{c}
P(t,\Delta T)=\int\widetilde{\Phi}(\overrightarrow{r};\overrightarrow{k},\Delta T)W\left(t,\left(T_{r}+T_{t}\right)\frac{c}{2},\Delta T,0\right)\\
V_{R}^{2}(\overrightarrow{r}_{SF})D_{\Sigma}(\overrightarrow{\Lambda}_{r},\overrightarrow{\Lambda}_{t},\overrightarrow{\Lambda}_{r},\overrightarrow{\Lambda}_{t},\omega_{0})\left|\gamma_{3}\right|^{2}F_{4}^{-1/2}(\overrightarrow{r}(\overrightarrow{k}),\overrightarrow{r}(\overrightarrow{k}))d\overrightarrow{k}\\
\\
\end{array}\label{eq:RLU-result-3D-spaceK}
\end{equation}

From qualitative considerations, the effective scattering volume $V_{R}$
should be close to that obtained for refraction-free case \cite{Berngardt_2000},
but instead of product $R_{t}R_{s}$ there should be the curvature
radii (as previously shown, they will automatically appear in the
two-dimensional stationary phase method over the angles (\ref{eq:WR_RgRg})):

\begin{equation}
V_{R} \approx \left\{ R_{GM,t}R_{Gm,t}R_{GM,r}R_{Gm,r}\right\} ^{1/2}k_{0}^{-3/2}(\overrightarrow{P}_{t}\times\overrightarrow{P}_{r})^{-1}\left(R_{t}+R_{r}\right)^{-1/2}\label{eq:WR-3D}
\end{equation}

The effective scattering volume $V_{R}$ is determined as:

\begin{equation}
V_{R}=\left|\aleph\right|^{-1/2}\label{eq:WR_def}
\end{equation}

where 
\begin{equation}
\aleph=\left\{ \frac{\partial}{\partial r_{j}}\left\{ k_{0}P_{t,i}+k_{0}P_{r,i}+k_{i}\right\} \right\} _{\overrightarrow{r}=\overrightarrow{r}_{0}}\label{eq:N_values}
\end{equation}

It is necessary to note that this condition corresponds to locally-homogeneous
trajectories. As can be seen from Fig.\ref{fig:2}, in the E-layer
(heights 100-130 km) both trajectories (transmitter-scatterer and
scatterer-receiver) can be considered as locally-homogeneous ones,
and correspondingly one can use a local formulation of the Wolf-Bragg
conditions (\ref{eq:WD-conditon-2st}). For the rest of the points,
in which the trajectory of locally-inhomogeneous, this approach is
not entirely correct and should be investigated separately. Expectedly,
that in this case the effective scattering volume will be less powerful,
and will be determined by the differential of the refraction index
along the propagation trajectory (similar to the case of the monostatic
scattering).

It is easy to show that in a refraction-free case the (\ref{eq:WR-3D})
transforms to the formulas obtained in \cite{Berngardt_2000}.

From the formulas (\ref{eq:WD-conditon-2st},\ref{eq:N_values}) it
is obvious that 
\begin{equation}
\aleph=\frac{\partial k_{i}(\overrightarrow{r})}{\partial r_{j}}\label{eq:JacCorrespondency}
\end{equation}
are the Lam\'{e} coefficients for the transition to the coordinate system
associated with the wave vectors $\overrightarrow{k}$ from the coordinate
system associated with spatial vectors of stationary points $\overrightarrow{r}$
(in the case that this transformation is bijection).

This means that in case of bijection $\overrightarrow{k}\rightarrow\overrightarrow{r}$
the integral over stationary points in the wave vectors space (\ref{eq:RLU-result-3D-spaceK})
can be represented as an integral over space $\overrightarrow{r}$
, where the exact expression for $V_{R}$ (\ref{eq:WR-3D}) is not
important:

\begin{equation}
\begin{array}{c}
P(t,\Delta T)=\int\widetilde{\Phi}(\overrightarrow{r},t-T_{t};\overrightarrow{k}(\overrightarrow{r}),\Delta T)W\left(t,\left(T_{r}+T_{t}\right)c,\Delta T,0\right)\\
\left|\gamma_{3}\right|^{2}D_{\Sigma}(\overrightarrow{\Lambda}_{r},\overrightarrow{\Lambda}_{t},\overrightarrow{\Lambda}_{r},\overrightarrow{\Lambda}_{t},\omega_{0})F_{4}^{-1/2}(\overrightarrow{r},\overrightarrow{r})d\overrightarrow{r}
\end{array}\label{eq:RLU-result-3D-spaceR}
\end{equation}

In this expression explicitly highlighted the dependence of the power
of the scattered signal from the scattering angle (\ref{eq:WD-conditon-2st})(in the refraction-free
case it is determined by the distance between the transmitter and
receiver) and from the Gaussian curvature of the beams $F_{4}^{1/2}$ (in the refraction-free
case it is determined by the distance to the investigated volume from
the receiver and transmitter, respectively). So, the effective weight
volume $V_{R}$ is defined by Jacobian for the transition from the wave vector
$\overrightarrow{k}$ space to usual space $\overrightarrow{r}$,
and is not important for spatial integral representation (\ref{eq:RLU-result-3D-spaceR}).
The shape of $V_{R}$ is important when one estimate the homogenety
scale for the media, necessary for the obtained formulas to be valid.
Also its exact shape should be important when the scattering is produced
by a single spatial harmonic (for example, during radioacoustical sounding
of the ionosphere \cite{Lataitis_1992}) and wavevector representation (\ref{eq:RLU-result-3D-spaceK}) 
may be more convenient.

\subsection{Multiple trajectories pairs}

If there are exist multiple propagation trajectories from the transmitter
to the scatterer ($M$ trajectories) and multiple paths from the scatterer
to the receiver ($N$ trajectories) the radar equation becomes more
complex than (\ref{eq:RLU-result-3D-spaceR}).

In this case integration over $\overrightarrow{\Lambda}_{t},\overrightarrow{\Lambda}_{t}'$
in (\ref{eq:RLU-result-3D-3-1}) will remove the delta function, but
produce some cross-sums over the trajectories, replacing the function
arguments with the corresponding functions $\overrightarrow{\Lambda}_{t,m}(\overrightarrow{r}),\overrightarrow{\Lambda}_{t,m'}'(\overrightarrow{r}'),\overrightarrow{\Lambda}_{r,n}(\overrightarrow{r}),\overrightarrow{\Lambda}_{r,n'}'(\overrightarrow{r}')$
(calculated based on the delta-function arguments), where indexes
$n,n',m,m'$ mark trajectories :

\begin{equation}
\begin{array}{c}
P(t,\Delta T)=\sum\limits _{n,n'\in N,m,m'\in M}\int\widetilde{\Phi}(\overrightarrow{r},t-T_{t,m};\overrightarrow{k},\Delta T+T_{t,m}-T_{t,m'}')\cdot\\
\cdot W\left(t,\left(T_{r,n}+T_{t,m}\right)c,\Delta T,\left(\left(T_{r,n}+T_{t,m}\right)-\left(T_{r,n'}'+T_{t,m'}'\right)\right)c\right)D_{\Sigma}(\overrightarrow{\Lambda}_{r,n},\overrightarrow{\Lambda}_{t,m},\overrightarrow{\Lambda}_{r,n'}',\overrightarrow{\Lambda}_{t,m'}',\omega_{0})\cdot\\
\cdot F_{4,n,n',m,m'}^{-1/2}(\overrightarrow{r},\overrightarrow{r'})\cdot e^{i\overrightarrow{k}\left(\overrightarrow{r}-\overrightarrow{r}'\right)}e^{ik_{0}\Psi_{n,m}(\overrightarrow{\Lambda}_{r,n},\overrightarrow{\Lambda}_{t,m})}e^{-ik_{0}\Psi_{n',m'}(\overrightarrow{\Lambda}_{r,n'}',\overrightarrow{\Lambda}_{t,m'}')}d\overrightarrow{r}d\overrightarrow{r}'d\overrightarrow{k}
\end{array}\label{eq:RLU-result-3D-3-1-2}
\end{equation}

where 
\begin{equation}
F_{4,n,n',m,m'}(\overrightarrow{r},\overrightarrow{r'})=F(\overrightarrow{\Lambda}_{r,n}(\overrightarrow{r}))F(\overrightarrow{\Lambda}_{r,n'}'(\overrightarrow{r}'))F(\overrightarrow{\Lambda}_{t,m}(\overrightarrow{r}))F(\overrightarrow{\Lambda}_{t,m'}'(\overrightarrow{r}'))
\end{equation}

In this case the condition of the stationary phase for each pair of
trajectories will take an already discussed form (\ref{eq:WD-conditon-2st}):

\begin{equation}
\overrightarrow{k}+k_{0}\overrightarrow{P}_{t}(\overrightarrow{\Lambda}_{t,m})+k_{0}\overrightarrow{P}_{r}(\overrightarrow{\Lambda}_{r,m})=0\label{eq:Wolf-Bragg-equation-bistatic}
\end{equation}

Assuming the stationary point $\overrightarrow{r}_{n,m}^{0}$ to be
only one for each given $\overrightarrow{k}$ and for each selected
unordered pair of trajectories $(n,m)$ the radar equation will take
the form:

\begin{equation}
\begin{array}{c}
P(t,\Delta T)=\sum\limits _{n,n'\in N,m,m'\in M,}\int\widetilde{\Phi}(\overrightarrow{r},t-T_{t,m};\overrightarrow{k},\Delta T+T_{t,m}-T_{t,m'}')\cdot\\
\cdot W\left(t,\left(T_{r,n}+T_{t,m}\right)c,\Delta T,\left(\left(T_{r,n}+T_{t,m}\right)-\left(T_{r,n'}'+T_{t,m'}'\right)\right)c\right)D_{\Sigma}(\overrightarrow{\Lambda}_{r,n},\overrightarrow{\Lambda}_{t,m},\overrightarrow{\Lambda}_{r,n'}',\overrightarrow{\Lambda}_{t,m'}',\omega_{0})\cdot\\
\cdot\left|\gamma_{3}\right|^{2}F_{4,n,n',m,m'}^{-1}(\overrightarrow{r}(\overrightarrow{k}),\overrightarrow{r'}(\overrightarrow{k}))\\
\cdot e^{i\overrightarrow{k}\left(\overrightarrow{r}_{n,m}^{0}-\overrightarrow{r}_{n',m'}^{0}\right)}e^{ik_{0}\Psi(\overrightarrow{\Lambda}_{r,n,m},\overrightarrow{\Lambda}_{t,n,m})}e^{-ik_{0}\Psi(\overrightarrow{\Lambda}_{r,n',m'}',\overrightarrow{\Lambda}_{t,n',m'}')}V_{R,n,m}^{2}d\overrightarrow{k}
\end{array}\label{eq:RLU-result-3D-several-traj}
\end{equation}

When transforming to the space coordinate system the effective weight
volume \textbf{$V_{R,n,m}$} disappears, because it is equal to the
Jacobian of the transition from one coordinate system ($\overrightarrow{k}$)
to another ($\overrightarrow{r}$) (\ref{eq:WR_def},\ref{eq:JacCorrespondency}):

\begin{equation}
\begin{array}{c}
P(t,\Delta T)=\sum\limits _{n,n'\in N,m,m'\in M}\int\widetilde{\Phi}(\overrightarrow{r},t-T_{t,m};\overrightarrow{k},\Delta T+T_{t,m}-T_{t,m'}')\cdot\\
\cdot W\left(t,\left(T_{r,n}+T_{t,m}\right)c,\Delta T,\left(\left(T_{r,n}+T_{t,m}\right)-\left(T_{r,n'}'+T_{t,m'}'\right)\right)c\right)D_{\Sigma}(\overrightarrow{\Lambda}_{r,n},\overrightarrow{\Lambda}_{t,m},\overrightarrow{\Lambda}_{r,n'}',\overrightarrow{\Lambda}_{t,m'}',\omega_{0})\cdot\\
\cdot\left|\gamma_{3}\right|^{2}F_{4,n,n',m,m'}^{-1/2}(\overrightarrow{r},\overrightarrow{r})e^{i\overrightarrow{k}\left(\overrightarrow{r}_{n,m}^{0}-\overrightarrow{r}_{n',m'}^{0}\right)}e^{ik_{0}\Psi(\overrightarrow{\Lambda}_{r,n,m},\overrightarrow{\Lambda}_{t,n,m})}e^{-ik_{0}\Psi(\overrightarrow{\Lambda}_{r,n',m'}',\overrightarrow{\Lambda}_{t,n',m'}')}d\overrightarrow{r}\\
\\
\end{array}\label{eq:RLU-result-3D-several-traj-1}
\end{equation}

\section{Resulting SuperDARN radar equation}

\begin{figure}
\includegraphics[scale=0.18]{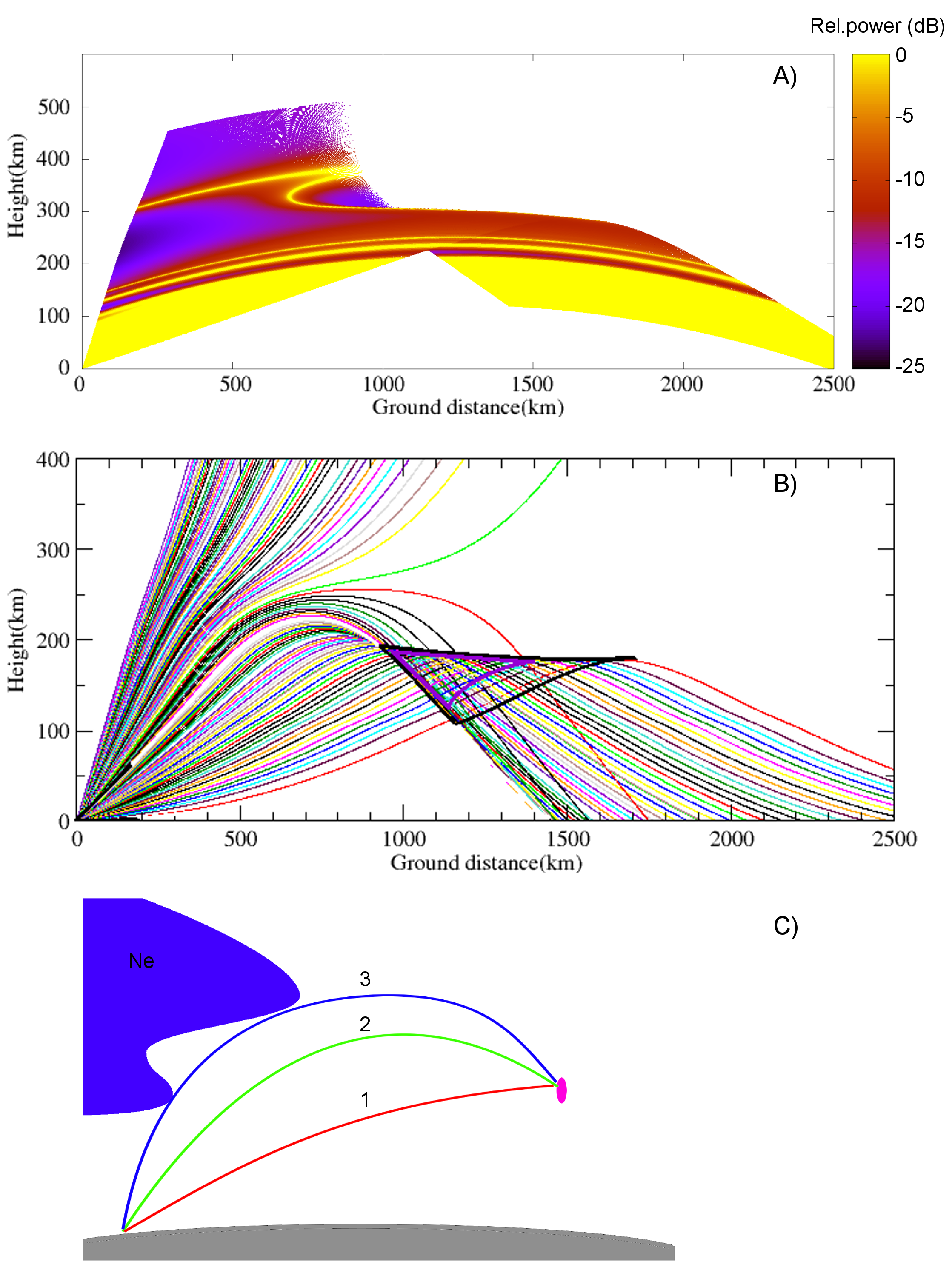} \protect\protect\protect\caption{Propagation effects in the radar equation. A) Effect of refraction
to the scattered signal power ($\left(J_{1}\right/\Delta R_{A})^{2}$
relation). Yellow areas corresponds to the scattering at locally-homogeneous
trajectories, red and blue - at locally-inhomogeneous trajectories.
B) The trajectories of the ray propagation. Multiple trajectory case
region is marked with a black triangle. Region with 3 trajectories
to any scattering point is marked with blue triangle. C) The geometry
of the multiple trajectory sounding. }

\label{fig:3} 
\end{figure}

In Fig.\ref{fig:3}B are shown the geometrooptical trajectories in
a model of spherically stratified ionosphere. From the figure it is
seen that within the region bounded by the caustics, there is a region
in which it is possible to reach the same point by multiple trajectories
(In Fig.\ref{fig:3}B the region is limited by the black triangle).
The upper limit for this region is the caustic associated with reflection
from the ionosphere by the lower ray, the left border is the caustic
associated with the boundary of the dead zone (i.e. ground backscatter
position), and the right border is the trajectory with the lowest
possible elevation angle.

Qualitative analysis has shown in the simple model case that the number
of trajectories to the each point, lying inside this region can be
two or three: the first is direct propagation trajectory, the second
is the trajectory reflected from the the ionosphere by lower ray,
and the last one is the trajectory reflected from the ionosphere by
the upper ray. Thus, in this area the case is possible when the signal
propagates towards and backwards the scatterer by substantially different
trajectories. The limit case for this effect is the area of caustics,
where to the same point at the same time comes a large number of rays,
that leads to significant gain of the signal.

In this paper we do not consider the caustic regions, but the area
inside the area bounded by the caustics can be considered.

The region, where all the 3 rays exist and intersect is marked by
blue triangle in Fig.\ref{fig:3}B. The region where at least two
trajectories exists is marked by black triangle. So the scattering
problem in the region can be considered as equivalent of bistatic
sounding problem.

In this case of several possible propagation trajectories from the
transmitter to the scatterer a situation arises that is equivalent
to the situation of bistatic sounding (\ref{eq:RLU-result-3D-several-traj},\ref{eq:RLU-result-3D-several-traj-1}).
This case does not arise in the refraction-free case, since in refraction-free
case there is only single trajectory exists between two points.

Lets consider the case of multiple trajectories in case of monostatic
sounding that is shown in Fig.\ref{fig:3}C. It can be seen that on
the trajectories, allowing the reflection from the ionosphere (i.e.,
at distances above 0.5 hop) there can be a few (at least two) signal
propagation trajectories to the scattering point. The trajectory marked
with 1 is the path of direct propagation (reflection-free trajectory).
The paths marked by 2,3 are the trajectories that includes reflection
from the ionosphere.

Lets obtain the radar equation in the case when all three trajectories
exists. When analyzing the signal autocorrelation function it is convenient
to analyze the simpliest case, valid for SuperDARN radars.

It is the case when selection over the delays is made with short enough
sounding pulses, spaced by large enough delay between each other.
The pulses are short enough to differ two trajectories by delay: $\left(T_{r,n}+T_{t,m}\right)-\left(T_{r,n'}'+T_{t,m'}'\right)>\Delta T_{A}$
where unordered pairs $(n,m)\neq(n',m')$. This condition also compensate
the additional caustic phase, that arise when wave reflects from the
ionosphere, and the caustic phase does not affect on resulting radar
equation.

In this case the short sounding pulse provides selection of correlating
trajectories when the difference between propagation delays is larger
than pulse duration. The final radar equation, after taking into account
symmetry considerations, becomes:

\begin{equation}
\begin{array}[t]{l}
\widetilde{P}(t,\omega)=\sum\limits _{n\in N}\int\limits _{|J_{1}/\Delta R_{A}|>1}\widetilde{\widetilde{\Phi}}\left(\overrightarrow{r};-\widehat{P}_{n}^{H}(\overrightarrow{r})k,\nu\right)A_{n}^{H}(\overrightarrow{r})\cdot\\
\,\,\cdot\widetilde{\widetilde{W}}_{2,t}\left(\hat{\Lambda}_{n}^{H}(\overrightarrow{r}),S_{n}(\overrightarrow{r}),\omega-\nu,\left(kP(\overrightarrow{r},\omega_{0})-2k_{0}P^{2}(\overrightarrow{r},\omega_{0})\right)-\nu/c\right)k^{2}dkd\nu d\overrightarrow{r}+\\
\\
+\int\limits _{|J_{1}/\Delta R_{A}|<1}\widetilde{\widetilde{\Phi}}\left(\overrightarrow{R}(\hat{\Lambda},S_{SF}(\hat{k} _{SF}^{I}k));\hat{k} _{SF}^{I}k,\nu\right)A^{I}(\overrightarrow{R}(\hat{\Lambda},S_{SF}(\hat{k} _{SF}^{I}k)))\cdot\\
\,\cdot\widetilde{W}_{2,t}\left(\hat{\Lambda},S_{SF}(\hat{k} _{SF}^{I}k),\omega-\nu,0\right)k^{2}dkd\nu d\hat{\Lambda}+\\
\\
+\sum\limits _{n,m\in N,n>m}\int\widetilde{\widetilde{\Phi}}(\overrightarrow{r};\overrightarrow{k}_{n,m}(\overrightarrow{r}),\nu)2(1+cos(\nu(T_{m}(\overrightarrow{r})-T_{n}(\overrightarrow{r}))))\cdot\\
\cdot A_{n,m}(\overrightarrow{r})\widetilde{W}\left(t,c(T_{n}(\overrightarrow{r})+T_{m}(\overrightarrow{r})),\omega-\nu,0\right)d\nu d\overrightarrow{r}
\end{array}\label{eq:final-SD-RLU3}
\end{equation}

where: 
\begin{equation}
\begin{array}[t]{l}
A_{n}^{H}(\overrightarrow{r})=\left(2\pi\right)^{2} k_{0}^{4} \left\{ \frac{2}{PF}J_{\hat{P}\rightarrow\hat{\Lambda}}\right\} \\
A^{I}(\overrightarrow{r})=\left(2\pi\right)^{3} k_{0}^{4} \left|g_{r}^{*}(\widehat{\Lambda},\omega_{0})g_{t}(\widehat{\Lambda},\omega_{0})\right|^{2}J_{1}^{2}J_{\hat{P}\rightarrow\hat{\Lambda}}\\
A_{n,m}(\overrightarrow{r})=\frac{\left(2\pi\right)^{3} k_{0}^{4}}{F_{n}F_{m}}\left|g_{r}^{*}(\widehat{\Lambda}_{n}(\overrightarrow{r}),\omega_{0})g_{t}(\widehat{\Lambda}_{m}(\overrightarrow{r}),\omega_{0})\right|^{2}
\end{array}
\end{equation}
are the multipliers that strongly affected by background ionosphere
and define power changes due to geometrooptical and focusing effects.

Here all the obvious arguments are hidden, and $J_{1},J_{\hat{P}\rightarrow\hat{\Lambda}},\widetilde{\widetilde{W}}_{2,t},\widetilde{W},F$
and $V_{\Lambda}$ are defined by (\ref{eq:J1},\ref{eq:J_P_L},\ref{eq:W_2t},\ref{eq:weight-volume-def},\ref{eq:dR_dL},\ref{eq:WR_RgRg})
correspondingly. $N$ is the maximal number of trajectories to reach
the same point at the same time. In analyzed case $N=[1,2,3]$: reflection-free
ray{[}1{]}, lower ray{[}2{]} and Pedersen ray{[}3{]} (see Fig.\ref{fig:3}C).

The radar equation (\ref{eq:final-SD-RLU3}) does not contain any
oscillating terms and is convenient for qualiative analysis or numerical
simulations of scattering process. In most cases, $\Phi(k)$ can be considered 
a smooth function of $k$ at scale $\Delta \omega / (cP)$  and radar equation (\ref{eq:final-SD-RLU3}) can be simplified even more.

The first two terms in (\ref{eq:final-SD-RLU3}) are associated with
the propagating of the signal over the identical trajectories towards
the scatterer and backwards. For any given scattering point whether
first or second term will be valid, because the integration of the
first term is made over the spatial area where $|J_{1}/\Delta R_{A}|>1$
(locally-homogeneous trajectory part), and second - over the spatial
area with $|J_{1}/\Delta R_{A}|<1$ (locally-inhomogeneous trajectory
part).

\begin{figure}
\includegraphics[scale=0.25]{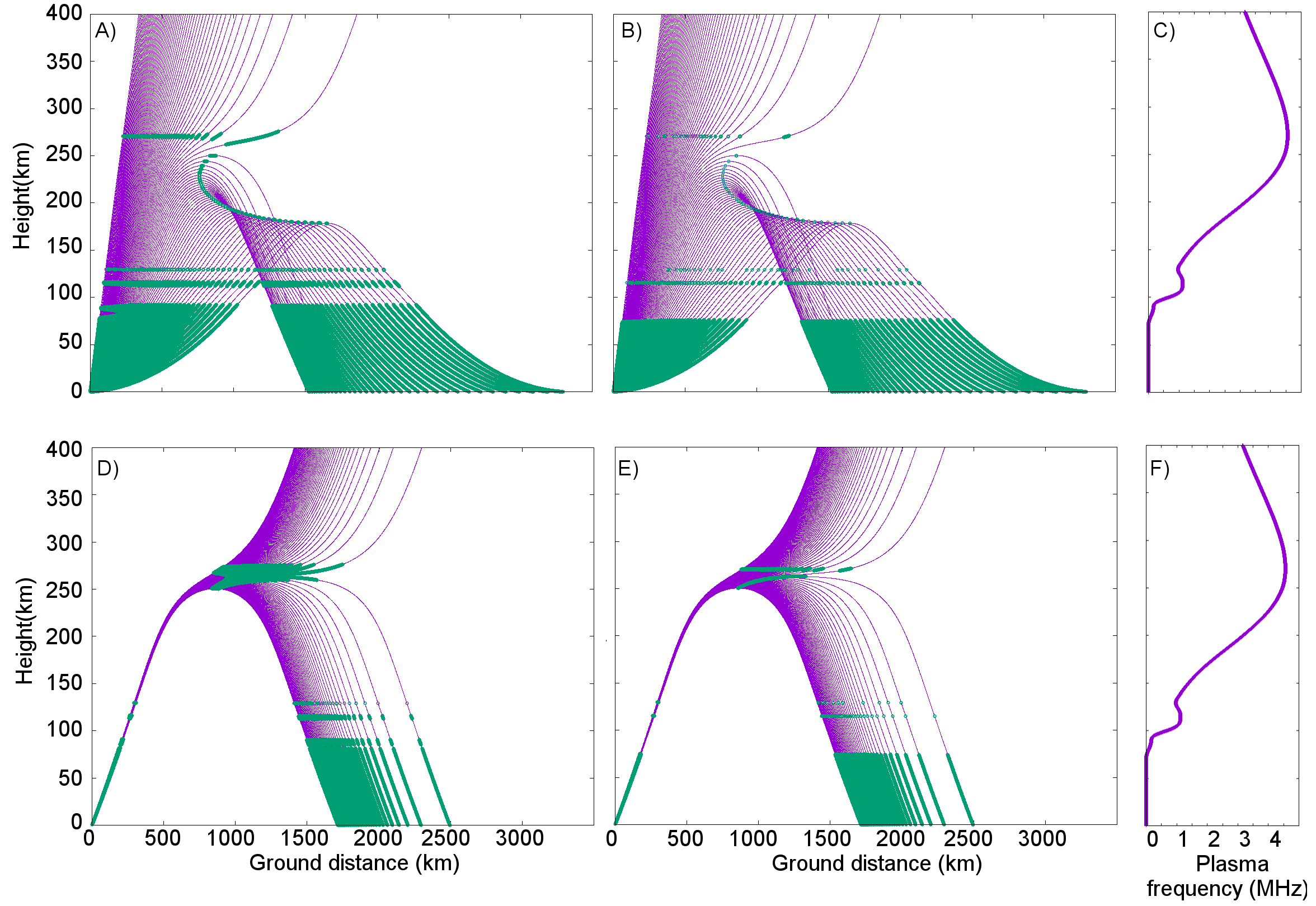} \protect\protect\protect\caption{ The areas of locally-homogeneous trajectories in the ionosphere.
Green areas corresponds to locally-homogeneous cases. A,D) for 100usec
sounding pulse, B,E) for 300usec sounding pulse, C,F) plasma frequency
profile used in the calculations. D-E) - the detailed areas near the
Pedersen ray.}

\label{fig:2} 
\end{figure}

The first term in the radar equation (\ref{eq:final-SD-RLU3}) corresponds
to the stationary points located at locally-homogeneous trajectories
(which is traditionally investigated in SuperDARN tasks or in refraction-free
case by VHF radars), the stationary point is marked by $H$ upper
index and is defined by (\ref{eq:ST_loc}). It's usage is limited
by locally-homogeneous condition $|J_{1}/\Delta R_{A}|>1$, for SuperDARN
case it is valid for very limited regions, including ionospheric E-layer
and F-layer maximums (see Fig.\ref{fig:2}). Exit angle $\hat{\Lambda}_{n}^{H}(\overrightarrow{r})$,
group path $S_{n}(\overrightarrow{r})$ and wave direction $\widehat{P}_{n}^{H}(\overrightarrow{r})$,
as well as $A_{n}^{H}(\overrightarrow{r})$ value, that correspond
to investigated point $\overrightarrow{r}$ can be calculated using
ray-tracing technique (\ref{eq:gam_eq_solution}). In refraction-free
case this term transforms to the refraction-free radar equation \cite{Berngardt_1999}.

The second term in the radar equation (\ref{eq:final-SD-RLU3}) corresponds
to the stationary points located at locally-inhomogeneous trajectories,
the stationary point is marked by $I$ upper index and is defined
by (\ref{eq:StatConditionScalar-LoclaKriv}). It's usage is limited
by locally-inhomogeneous condition $|J_{1}/\Delta R_{A}|<1$. For
SuperDARN case it is valid for most part of the F-layer ionosphere
(see Fig.\ref{fig:2}). This case has no equivalent in refraction-free
situation. To calculate it one need to find the point with most intensive
scattering $S_{SF}(\hat{\Lambda},k)$ that is defined by system (\ref{eq:StatConditionScalar-LoclaKriv}).
It looks complex task, but the simplest way to estimate it, is to
calculate the wavevector direction $\hat{k}_{SF}^{I}$ that in average
is perpendicular to wavefront according to the first condition in
(\ref{eq:StatConditionScalar-LoclaKriv}) for given exit angle $\hat{\Lambda}$
and over weight volume $W(..)$ size: $\int\limits _{ct/2-\Delta R_{A}}^{ct/2+\Delta R_{A}}\hat{k}_{SF}^{I}\left(\frac{\partial\overrightarrow{P}}{\partial\Lambda}\right)_{\bot\overrightarrow{P}}dS_{1}=0$
. Then one can numerically find the point $S_{SF}$, at which the
Wolf-Bragg condition (second condition in (\ref{eq:StatConditionScalar-LoclaKriv}),
$k\hat{k}_{SF}^{I}\overrightarrow{P}(\hat{\Lambda},S_{SF})-2k_{0}P^{2}(\hat{\Lambda},S_{SF})-\nu/c=0$)
is satisfied for given wavenumber $k$, exit angle $\hat{\Lambda}$
and calculated average wavefront perpendicular $\hat{k}_{SF}^{I}$. It
is important to note, that $S_{SF}$ should be $\in[ct/2-\Delta R_{A},ct/2+\Delta R_{A}]$,
or integral becomes neglectable because the carrier of weight volume
$W(...)$. The $A^{I}(\overrightarrow{r})$ value and group delay
$T_{n}(\overrightarrow{r})$ to the investigated point can be calculted
during ray-tracing.

The third term (cross-sums) in the radar equation (\ref{eq:final-SD-RLU3})
is associated with propagation towards the scatterer over one trajectory
and backwards to the receiver over another trajectory, and is an equivalent
of bistatic scattering\cite{Berngardt_2000}. The stationary point
is defined by (\ref{eq:Wolf-Bragg-equation-bistatic}). Wavevector
$\overrightarrow{k}_{n,m}(\overrightarrow{r})$ can be found using
Wolf-Bragg condition (\ref{eq:Wolf-Bragg-equation-bistatic}) by ray-tracing
technique (\ref{eq:gam_eq_solution}) over two different trajectories
$n,m$ to the investigated point $\overrightarrow{r}$. $A_{n,m}(\overrightarrow{r})$
value, group delays $T_{n}(\overrightarrow{r}),T_{m}(\overrightarrow{r})$
and exit angles $\Lambda_{n}(\overrightarrow{r}),\Lambda_{m}(\overrightarrow{r})$
to the investigated point can be also calculated during ray-tracing.
This case does not arise in refraction-free situation. It should be
noted that existence of the third term was considered and discussed
for a static scatterer in \cite{Kravcov_1980}.

To take into account a weak absorption $\chi$ during radiowave propagation,
for example in D-layer \cite{Ginzburg_1970,Settimi_2014,Settimi_2015},
one should include the multiplier $e^{-2\int_{0}^{cT_{n}}\chi(\overrightarrow{R}(\hat{\Lambda}_{n},S',\omega_{0}))dS'}$
into each of $A_{n}^{H}(\overrightarrow{r}),A^{I}(\overrightarrow{r})$
and the multiplier $e^{-\int_{0}^{cT_{n}}\chi(\overrightarrow{R}(\hat{\Lambda}_{n},S',\omega_{0}))dS'-\int_{0}^{cT_{m}}\chi(\overrightarrow{R}(\hat{\Lambda}_{m},S',\omega_{0}))dS'}$
into $A_{n,m}(\overrightarrow{r})$.

\section{Discussion and conclusion}

Based on hamiltonian optics approach and ray representation of Green's
function we obtained the scalar radar equation (\ref{eq:final-SD-RLU3})
that is suitable for taking into account the refraction effects in
SuperDARN data (autocorrelation functions of the received signals)
obtained by these radars within the first hope distances (i.e. before
scattering of the signal from the ground). The main conclusions from
this radar equation are the following.

1. The radar equation shows that the integral kernel that relates
the measured correlation function of the signal with the correlation
function of the ionospheric irregularities, is the weight volume (or
ambiguity function) $W(...)$(\ref{eq:weight-volume-def}), the widely
studied for refraction-free case \cite{Lehtinen_1986,Berngardt_1999,Berngardt_2000}.
This weight volume is defined by the shape of a sounding signal and
by signal processing techniques. 
Radar equation (\ref{eq:final-SD-RLU3}) is linear over $W(...)$, so making any linear transformation of the scattered signal correlation function or spectral power 
leads just to equivalent transformation of the weight volume $W(...)$ \cite{Lehtinen_1986,Berngardt_2013} and does not affect on the structure 
of the radar equation.
This allows, for example, to prove the validity
of using at SuperDARN radars the signal processing techniques and sounding sequences,
initially developed for refraction-free cases (\cite{Farley_1972,Berngardt_2015c},
etc.).

2. The radar equation shows that the spatial harmonics that produce
the main contribution to the scattering signal are determined by the
local wave vector and satisfy to the Wolf-Bragg condition (\ref{eq:ST_loc},\ref{eq:StatCondition-Local-mod}).
This is very close to refraction-free case \cite{Berngardt_1999,Berngardt_2000}.
This also confirms with the results of \cite{Gillies_2011}, and
allows to take it into account refraction effects in the velocities
of ionospheric irregularities measured by SuperDARN radars.

3. In contrast to the refraction-free case, the refraction significantly
affects the scattered signal amplitude. At first, this effect leads
to amplitude changes related to the focusing of the signal during
propagation (multipliers $F^{-1/2}$) in the region of the scattering.
The second effect is associated with attenuation of the scattered
signal due changing the wave length of sounding signal at intervals
of the length of the sounding pulse (parameter $J_{1}$ in the equation).

4. Obtained radar equation is not exactly correct for the Pedersen
rays after the point of theirs reflection. This is associated with
considerable frequency distortion of the propagating signal in this
area ( higher-frequency components are not reflected by the ionosphere)
and, as a consequence, the inapplicability of the undistorted signal
propagation model.

5. The highest amplitude of the scattered signal occurs in regions
close to the areas of focusing, and in the areas where the refractive
index varies slightly along the propagation trajectories. These areas
correspond to the caustic trajectories, the areas near the local maximum
of the electron density (Pedersen rays) and the areas where the ionospheric
refraction can be neglected (the areas below the E-layer). Outside
these areas, the amplitude of the scattered signal is reduced and
can be compensated only by a high aspect dependence of scattering
irregularities. The maximal scattering amplitude at Pedersen ray have
been detected, for example, as a result of interpretation of the first
Russian-Ukrainian experiment using EKB radar and UTR-2 radiotelescope
\cite{Berngardt_2015b}.

\begin{figure}
\includegraphics[scale=0.2]{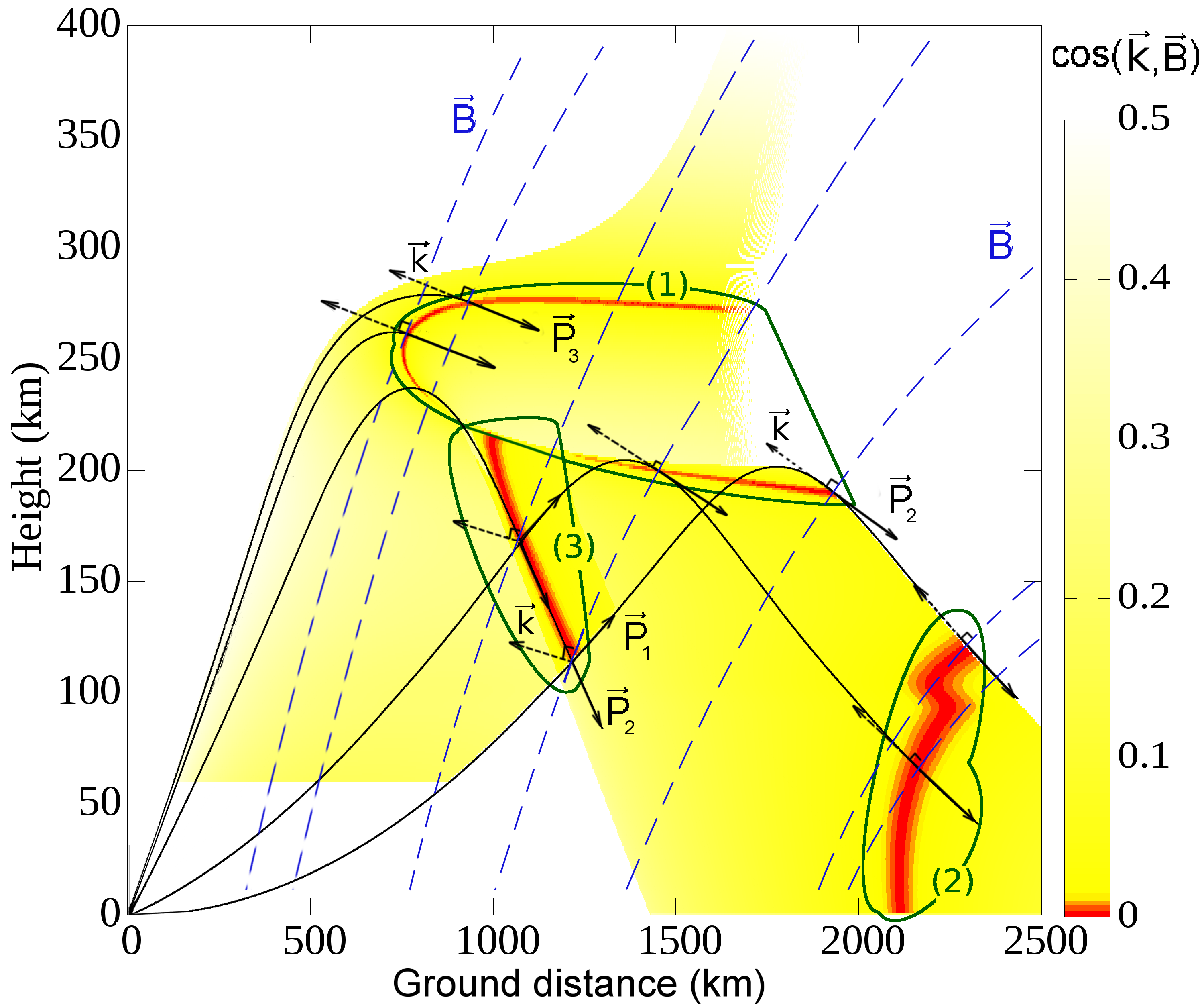}

\protect\protect\protect\caption{Model calculations for regions of most intensive scattering from field-aligned
irregularities for a radar, located at north magnetic pole and looking
to the equator (calculations are based on IGRF reference magnetic
field model and IRI-2012 reference ionosphere model). Black lines
- ray trajectories $R(\widehat{\Lambda},S)$; blue lines - magnetic
field $\protect\overrightarrow{B}$ lines ; green regions - orthogonality
regions for monostatic scattering at F-layer heights (1), for monostatic
scattering at E-layer heights (2) and for multi-trajectory scattering
(3). }

\label{fig:6} 
\end{figure}

6. When signal propagates in an inhomogeneous ionosphere there are
areas in which the signal reaches the point of scattering by several
(two or more) different trajectories simultaneously. We modeled the
orthogonality conditions of wave vector to the magnetic field, important
for scattering by field-aligned irregularities, with taking into account
the structure of the Earth magnetic field (IGRF reference model used)
and model ionosphere (IRI-2012 reference model used). Modeling shows
that the bistatic mechanism should be important for polar radars looking
equatorward (see Fig.\ref{fig:6}). In spite in this case the angle
between propagation trajectory and magnetic field is far from orthogonality (about
50 degrees), the wavevector of the scattering irregularities $\overrightarrow{k}$
is orthogonal to magnetic field $\overrightarrow{B}$. This mechanism
allows to interpret strong field-aligned scattering even at very high
aspect angles between trajectory and magnetic field.

7. The first two mechanisms can be considered as a result of monostatic
scattering similar to the refraction-free case \cite{Tatarsky_1967,Ishimaru_1999,Berngardt_1999}.
The third mechanism should be considered as an equivalent of bistatic
sounding in refraction-free case \cite{Tatarsky_1967,Ishimaru_1999,Berngardt_2000}. This case
apparently has not been considered in SuperDARN tasks. The existence
of a third mechanism has been proposed and investigated for the case
of stationary irregularities for example in \cite{Kravcov_1980}.
It should be noted that in the last case, despite the possibility
of differencing the trajectories over the time of arrival, the measured
spectrum is distorted by $2(1+cos(\nu(T_{m}-T_{n})))$ multiplier
due to trajectory pairs with exactly the same group length. Conversely,
the existence of such distortions in the received signal spectral power allows to
detect the multiple propagation case.

\section{acknowledgments} The work of OB and AP has been done under
support of FSR program II.12.2.2. The work of KK has been done under
support of RFBR grant \#16-05-01006a. The numerical simulations have
been performed with High-performance computing cluster "Academician
V.M. Matrosov" (http://hpc.icc.ru) 

\bibliography{ARTICLE}

\newpage

\end{document}